\documentclass[a4paper,12pt]{article}
\pdfoutput=1 

\usepackage{jheppub}

\usepackage[T1]{fontenc}
\usepackage[utf8]{inputenc}

\usepackage{amsmath}
\usepackage{graphicx}
\usepackage{xcolor}
\usepackage[normalem]{ulem}
\usepackage{amsthm}
\usepackage{bm}
\usepackage{physics}

\usepackage{hyperref}
\usepackage[capitalize]{cleveref}

\usepackage{subcaption}

\theoremstyle{definition}

\newcommand{\be}{\begin{equation}}
\newcommand{\ee}{\end{equation}}
\newcommand{\beal}{\begin{aligned}}
\newcommand{\eeal}{\end{aligned}}
\newcommand\bea{\begin{eqnarray}}
\newcommand\eea{\end{eqnarray}}
\newcommand{\bec}{\begin{cases}}
\newcommand{\eec}{\end{cases}}

\title{\boldmath Properties of black holes in non-linear electrodynamics}

\author[a]{Lewis Croney,}
\author[a,b]{Ruth Gregory,}
\author[a,1]{Ansh Gupta,\note{Corresponding author.}}
\author[a,c]{Carlos J. Ramírez-Valdez}

\affiliation[a]{Department of Physics, King’s College London, University of London, Strand, London, WC2R 2LS, UK}
\affiliation[b]{Perimeter Institute, 31 Caroline Street North, Waterloo, ON, N2L 2Y5, Canada}
\affiliation[c]{Escuela de Ingeniería y Ciencias, Instituto Tecnológico y de Estudios Superiores de Monterrey, Campus Estado de México, Estado de México, México}

\emailAdd{lewis.croney@kcl.ac.uk}
\emailAdd{ruth.gregory@kcl.ac.uk}
\emailAdd{ansh.gupta@kcl.ac.uk}
\emailAdd{jonathan.ramirez@tec.mx}

\abstract{We investigate the properties of charged black hole geometries in nonlinear electrodynamics. We focus on the recently reported analytic charged black hole solutions to illustrate the consequences of a non-monotonic lapse function that exists for a wide range of black hole solutions. The spacetime admits stable light-rings, static near-horizon observers, and trapped near horizon photon orbits. We also show that although these modifications near the horizon are screened from afar, they nonetheless lead to additional branches of quasinormal modes for the black hole that are longer lived than the canonical Einstein branches. }

\begin{document}
\maketitle
\flushbottom

\section{Introduction}
Black holes constitute one of the most fundamental predictions of Einstein’s General Relativity (GR), providing a natural laboratory for probing the interplay between gravitation, and high-energy physics. Within the classical Einstein--Maxwell framework, electrically and magnetically charged solutions such as the Reissner--Nordstr\"om spacetime are well understood. Nevertheless, this linear theory suffers from intrinsic limitations, most notably the divergence of the electromagnetic self-energy of point charges \cite{Born:1934,Heisenberg:1936} and the persistence of curvature singularities \cite{Hawking:1970}. These shortcomings strongly motivate the exploration of extensions to both the gravitational and electromagnetic sectors, particularly in contexts inspired by quantum field theory and string theory \cite{Wiltshire:1988uq,Born:1934,Heisenberg:1936,Fradkin:1985}.

Nonlinear electrodynamics (NLED) provides a natural extension of Maxwell’s theory by allowing the electromagnetic Lagrangian to depend nonlinearly on the field invariants
\begin{align}
F &= F_{\mu\nu}F^{\mu\nu}, \qquad
G = F_{\mu\nu}\tilde{F}^{\mu\nu}.
\end{align}
Originally introduced in the seminal work of Born and Infeld to regularise the divergent self-energy of point charges \cite{Born:1934}, NLED has since emerged as an effective description arising from string theory \cite{Fradkin:1985,Leigh:1991}, quantum electrodynamics corrections such as the Euler--Heisenberg model \cite{Heisenberg:1936}, and gauge/gravity duality \cite{Jin:2010, cai2004}. When coupled to GR, NLED can substantially modify spacetime geometry, giving rise to black hole solutions with novel causal, thermodynamic, and dynamical properties, including families of regular black holes free from curvature singularities \cite{Bardeen:1968,AyonBeato:1998,AyonBeato:1999,Bronnikov:2000, Verbin:2024ewl}, while further offering critical insights into holographic superconductors and the extended phase space thermodynamics of asymptotically anti-de Sitter (AdS) spacetimes \cite{Hartnoll:2008,Kastor:2009,Gunasekaran:2012,Kubiznak:2017}.

In linear Maxwell theory, charged black hole solutions exhibit a symmetry under the interchange of electric and magnetic charges. This duality symmetry is generically broken in more intricate theories, such as those involving additional fields \cite{Gibbons:1987}, holographic constructions \cite{Hartnoll:2007}, or nonlinear electromagnetic interactions themselves \cite{Bronnikov:2000}. This naturally motivates the study of \textit{dyonic} black holes, which carry both electric and magnetic charges. In NLED, such configurations display a much richer phenomenology, significantly altering the thermodynamic behaviour and phase structure relative to the Reissner--Nordstr\"om--AdS case.

Early investigations primarily explored black holes within specific NLED frameworks, most notably Born--Infeld electrodynamics \cite{Wiltshire:1988uq, Fernando:2003}. More recently, a complementary approach has gained prominence, wherein NLED is treated as a weak-field expansion in the electromagnetic invariants $F$ and $G$ \cite{Yajima:2001, Kruglov:2016ezw, Hendi:2013dwa}. However, deriving exact solutions within these expansions remains analytically challenging due to the intricate coupling and heightened complexity of the resulting field equations; consequently, research often focuses on obtaining solutions only to first order in the coupling parameters \cite{Yajima:2001}. Herein lies the importance of analytical solutions in the weak-field expansion regime, such as the solutions recently reported in \cite{Croney:2025}. These solutions allow the effects of nonlinear electrodynamics to be studied directly, without introducing additional perturbative corrections beyond those already present in the Lagrangian. This provides a setting in which to isolate some phenomena associated with nonlinearity. In particular, the emergence of non-monotonic metric functions in these regimes can lead to a double-barrier effective potential, fundamentally altering the stability of particle orbits \cite{Wang:2025wob}.

A significant aspect are the modifications introduced by NLED to the motion of test particles and light. In contrast to Maxwell theory, photons in NLED do not generally follow null geodesics of the background spacetime metric. Instead, they propagate along null geodesics of an effective geometry $g_{\mu\nu}^{\text{eff}}$ determined by the nonlinear electromagnetic interactions \cite{Boillat:1970,Novello:1999pg, DeLorenci:2000yh, GoulartdeOliveiraCosta:2009pr}. This results in the phenomenon of vacuum birefringence, where different polarisation modes experience distinct optical metrics \cite{Obukhov:2002xa, Ovgun:2025mdg}. Such nonlinearities can produce additional light-rings and stable trapping regions \cite{marks2025, fonseca2025}, which are often accompanied by acausal regions if the Maxwell weak-field limit is violated \cite{Murk:2024nod}. From the perspective of the background metric, this behaviour can be interpreted as a non-geodesic motion driven by an additional effective force. These effects systematically shift the photon-sphere radius and modify the critical impact parameter, producing apparent black hole shadows that can be up to $\mathcal{O}(10\%)$ larger than those predicted by linear electrodynamics \cite{dePaula:2023, Tang:2023lmr, Yuan:2024wdl, AraujoFilho:2024xhm}.

For timelike particles, NLED couplings deepen the effective potential well, increasing the binding energy of stable circular orbits \cite{Alloqulov:2025}. More sophisticated models, such as Euler--Heisenberg or ModMax electrodynamics, further predict phenomena including vacuum birefringence and the disappearance of near-horizon potential barriers, which may lead to qualitatively distinct perturbative regimes \cite{Fathi:2025}.

Beyond the polarisation of light, NLED profoundly affects black hole spectroscopy, i.e., quasinormal modes (QNMs) \cite{Berti:2009,Jansen:2017, QNMs_Review}. Analytic treatments of these spectra reveal that the eikonal limit provides a direct bridge between QNMs and the aforementioned null geodesic trapping \cite{Bolokhov:2025uxz}. Importantly, although NLED can resolve curvature singularities, recent studies \cite{Nomura:2021efi} have revealed that quasinormal spectra exhibit qualitative deviations tied to nonlinear effects, including the breaking of isospectrality under parity transformations \cite{Nomura:2021efi}. This underscores that dynamical analysis is essential for assessing the physical viability of NLED solutions. These geodesic and spectral deviations provide powerful observational diagnostics for constraining nonlinear couplings using data from the Event Horizon Telescope and gravitational-wave detectors \cite{Pantig:2022gih, Aliyan:2024xwl, Zare:2024dtf}.

This work explores the properties of the novel exact magnetic black hole solution reported in \cite{Croney:2025} within the framework of NLED. 
This solution is an exemplar for the effect of non-monotonic lapse function, with a rich geodesic structure, fundamentally distinct optical metrics, and new branches of quasinormal modes (QNMs). We first review analytic NLED solutions representative of the fluctuating lapse function in \S \ref{sec:dyonicbh}. Section \ref{sec:geodesics} is devoted to an analysis of geodesics: null, timelike, and optical in the spacetime, while \S \ref{sec:qnms} examines the QNM spectrum. Section \ref{sec:discussion} concludes.

\section{NLED black hole solution} \label{sec:dyonicbh}

We begin by reviewing the dyonic NLED black hole solution presented in \cite{Croney:2025}.

Our system has the following action,
\begin{equation}
\mathcal{S} = \frac{1}{16\pi} \int_{\mathcal{M}^4} d^4 x \sqrt{-g} (R - F + aF^2 +bG^2),
\label{action}
\end{equation}
where $R$ is the Ricci scalar, and $F$ and $G$ are electromagnetic invariants defined as $F:= F_{\mu\nu}F^{\mu\nu}$ and $G:= F_{\mu\nu}\tilde{F}^{\mu\nu}$.

The corresponding equations of motion for this action are given by,
\begin{align}
    R_{\mu\nu}-\frac{1}{2}g_{\mu\nu}R&=8\pi T_{\mu\nu},\\
        \nabla^\mu P_{\mu\nu}&=0, \label{eqn:P_eqn}
\end{align}
where the energy momentum tensor $T_{\mu\nu}$, and the auxiliary tensor $P_{\mu\nu}$ are given by,
\begin{align}   
    T_{\mu\nu}&=\frac{1}{4\pi}\left[(1-2aF)F_{\mu\alpha}F_{\nu}^{\;\alpha} - \frac{1}{4} g_{\mu\nu}(F-aF^2-bG^2)\right],\\
    P_{\mu\nu}&=F_{\mu\nu}-2aF F_{\mu\nu} - 2bG\tilde{F}_{\mu\nu}.
\end{align}

A static, spherically symmetric solution has a metric with the form,
\begin{equation}\label{eqn:metric}
    ds^2=-f(r)dt^2+\frac{1}{f(r)}dr^2+r^2(d\theta^2+\sin^2\theta d\varphi^2),
\end{equation}
where $f(r)$ can in principle be found from the schematic form,
\begin{equation} \label{eqn:f(r)full}
    f(r) = 1 - \frac{2m}{r}  + \frac{p^2}{r^2} -\frac{2ap^4}{5r^6}+ \frac{1}{2r} \int^r dr' \left[\left(r'^2 + \frac{4p^2}{r'^2}(2b-a) \right) F_{rt}^2 - 3q F_{rt} \right].
\end{equation}
Here, $m$ is a constant of integration which is usually associated with the black hole ADM mass, $p$ is the (topological) magnetic charge in the Maxwell field, $F_{\theta\phi}=p\sin\theta$, $q$ is a constant arising in the solution $P_{rt}=q/r^2$ to \eqref{eqn:P_eqn} corresponding to the electric charge as measured at infinity, and $F_{rt}$ can be determined by solving the cubic,
\begin{equation}
4 a F_{rt}^3 + \left(1 + \frac{4p^2}{r^4}(2b-a) \right) F_{rt} - \frac{q}{r^2} = 0.
\label{Fcubic}
\end{equation}
We can use this expression to rewrite the integral in \eqref{eqn:f(r)full} as
\be
\int dr' \left[\left(r'^2 + \frac{4p^2}{r'^2}(2b-a) \right) F_{rt}^2 - 3q F_{rt} \right] = -
\int dr' \left[ 4 a r'^2 F_{rt}^4 + 2q F_{rt} \right].
\ee
Note that this integrand is invariant under $q \leftrightarrow -q$, and is strictly positive, therefore setting the integral limits to have the correct limit of $f$ at infinity, we obtain
\be
f(r) = 1 - \frac{2m}{r} + \frac{p^2}{r^2} -\frac{2ap^4}{5r^6}+ \frac{1}{r} \int_r^\infty dr' \left[ 2 a r'^2 F_{rt}^4 + q F_{rt} \right].
\label{fofr}
\ee

In spite of the apparent complexity of the expressions for $F_{rt}$ and $f(r)$, we are aware of two special cases where $f(r)$ has an analytic form: 
The magnetic black hole reported in \cite{Croney:2025}, and a dyonically charged black hole with $a=0$ reported in \cite{Wang:2025wob}. 
While we have been unable to reproduce the expression $f(r)$ from \cite{Wang:2025wob}, we confirm the existence of lapse functions $f(r)$ with qualitatively the same behaviour in \S \ref{subsec:quasitop}, presenting the analytic solution to the Einstein equations with $a=0$. 
We now briefly note these analytic solutions.

\subsection{The magnetic black hole}

Setting $q=0$, the asymptotically flat, purely magnetic black hole has a relatively simple analytic form of the metric \cite{Croney:2025}:
\begin{align} \label{eqn:f(r)}
f(r) = 1 - \frac{2m}{r}  + \frac{p^2}{r^2} -\frac{2ap^4}{5r^6} = f(r)_{\text{RN}} - \frac{2ap^4}{5r^6}.
\end{align} 
In the linear Reissner-Nordström (RN) case there is an extremal limit as $p\rightarrow m$, leading to a naked singularity for $p>m$. However, there is no such limit in our case as a horizon is guaranteed for all $m$. Instead, parameters may be restricted due to chronology 
or causality constraints, which can be shown to be equivalent to energy 
constraints \cite{Russo:2024xnh, Russo:2026vnj}. Imposing the dominant energy condition, $\rho \geq |p_i|$, where
\be
\rho = -p_r = p^2/r^4 - 2ap^4/r^8\;,\qquad
p_\theta = p_\phi = p^2/r^4 - 6ap^4/r^8,
\ee
(implying $r_h^4>4ap^2$) avoids pathologies, see discussions in \cite{Babichev:2007dw,Croney:2025}.

\begin{figure}[h]
\begin{subfigure}{0.49\textwidth}
    \includegraphics[width=\textwidth]{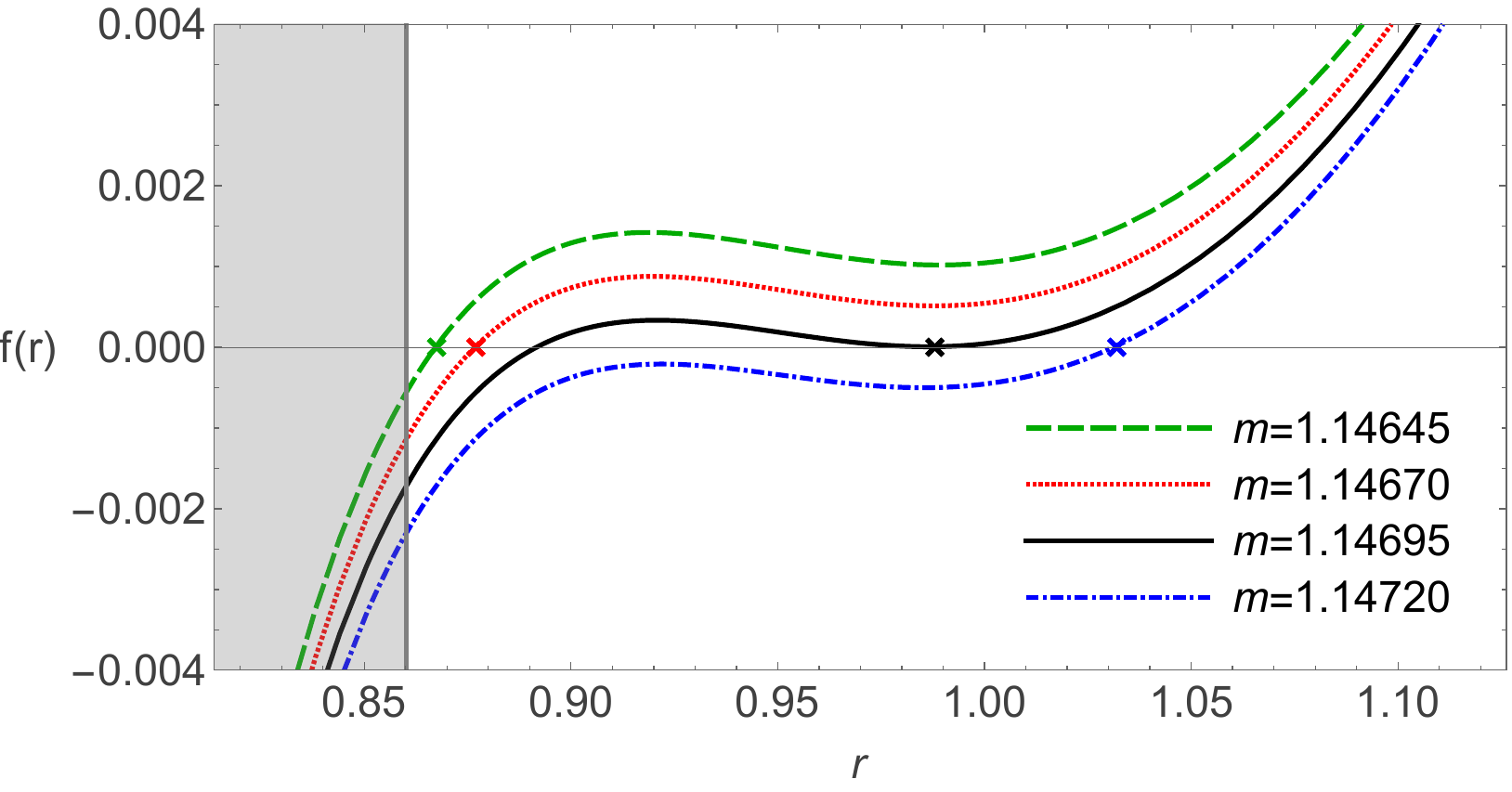}
\end{subfigure}
\begin{subfigure}{0.49\textwidth}
    \includegraphics[width=\textwidth]{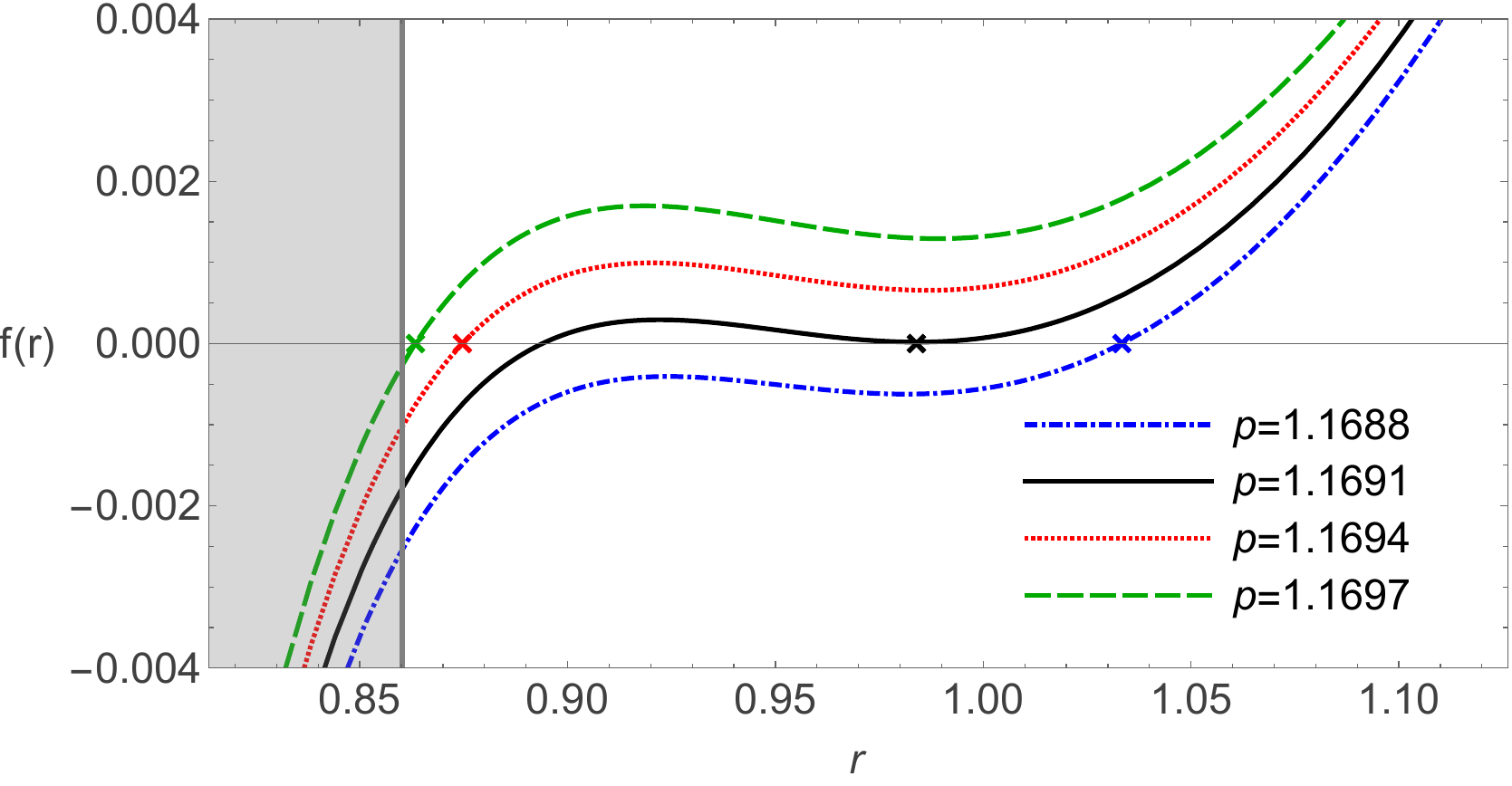}
\end{subfigure}
\caption{A plot of $f(r)$ showing a ``jumping'' event horizon for a magnetic black hole as either the mass or the magnetic charge is varied. \textbf{Left:} The magnetic charge is fixed, $p=1.17$, while $m$ is varied. \textbf{Right:} The mass is fixed, $m=1.146$, while $p$ is varied. In both plots, the vertical line marks the threshold for Dominant Energy Condition (DEC) violation, so each curve shown conserves DEC outside the horizon. The black hole event horizon in each case is given by the largest zero-crossing of $f(r)$, and denoted by a cross. As the minimum in $f(r)$ dips below 0, there is a sudden jump in horizon radius. This figure is similar to that in \cite{Croney:2025}, with parameters $q=0, a=0.1$.}
\label{fig:f(r)jump}
\end{figure}

Tuning the parameters of our system produces the ``jumping horizon'' behaviour first seen in \cite{Croney:2025}, and reproduced in figure \ref{fig:f(r)jump}. It is this region of parameter space that we primarily investigate in this paper, since we expect a non-monotonic $f(r)$ to result in interesting behaviour of geodesics and quasinormal modes (QNMs).

\subsection{The quasi-topological black hole}
\label{subsec:quasitop}

Setting $a=0$ in \eqref{action} gives a quasi-topological electromagnetic Lagrangian that can be rewritten as
\be
-F + b G^2 = -F + b \left [4 F^\mu_{~\nu} F^\nu_{~\sigma} F^\sigma_{~\lambda} F^\lambda_{~\mu} - 2 \left ( F_{\mu\nu}F^{\mu\nu} \right)^2 \right ],
\ee
(up to rescalings this is the Lagrangian considered in \cite{Wang:2025wob}). 
With $a=0$, \eqref{Fcubic} ceases to be a cubic and gives
\be
F_{rt} = \frac{q r^2}{r^4+8bp^2}
\ee
Writing $\alpha^4 = 2 b p^2$ to simplify notation, this leads to the analytic expression
\be
\beal
f_D(r) = & 1 - \frac{2m}{r} + \frac{p^2}{r^2} + \frac{q^2}{8r\alpha} \Biggl [
2\pi + \log \left ( \frac{r^2 + 2 \alpha r + 2 \alpha^2}{r^2 - 2 \alpha r + 2 \alpha^2} \right )\\
& \hskip 2cm
- 2 \arctan\left ( \frac{r+\alpha}{\alpha} \right )- 2 \arctan\left ( \frac{r-\alpha}{\alpha} \right )
\Biggr ]
\eeal
\label{dyonicf}
\ee
where we have added the subscript $D$ to denote this as a dyonic solution.
As with the magnetic black hole, for a finely tuned set of parameters this lapse function also has turning points outside the horizon, and the possibility of a horizon jump for infinitesimal increments of mass. 
In distinction with the magnetic black hole however, the dominant energy condition for this solution is always satisfied:
\be
\begin{matrix}
\rho = - p_r = \left ( 1 + \frac{8bp^2}{r^4} \right ) F_{rt}^2 + \frac{p^2}{r^4}\\
p_\theta = p_\phi = \left ( 1 - \frac{8bp^2}{r^4} \right ) F_{rt}^2 + \frac{p^2}{r^4}
\end{matrix}
\quad \Biggr \}
\Rightarrow \rho \pm p_i \geq 0
\ee

\begin{figure}[h]
\centering
\includegraphics[width=0.8\linewidth]{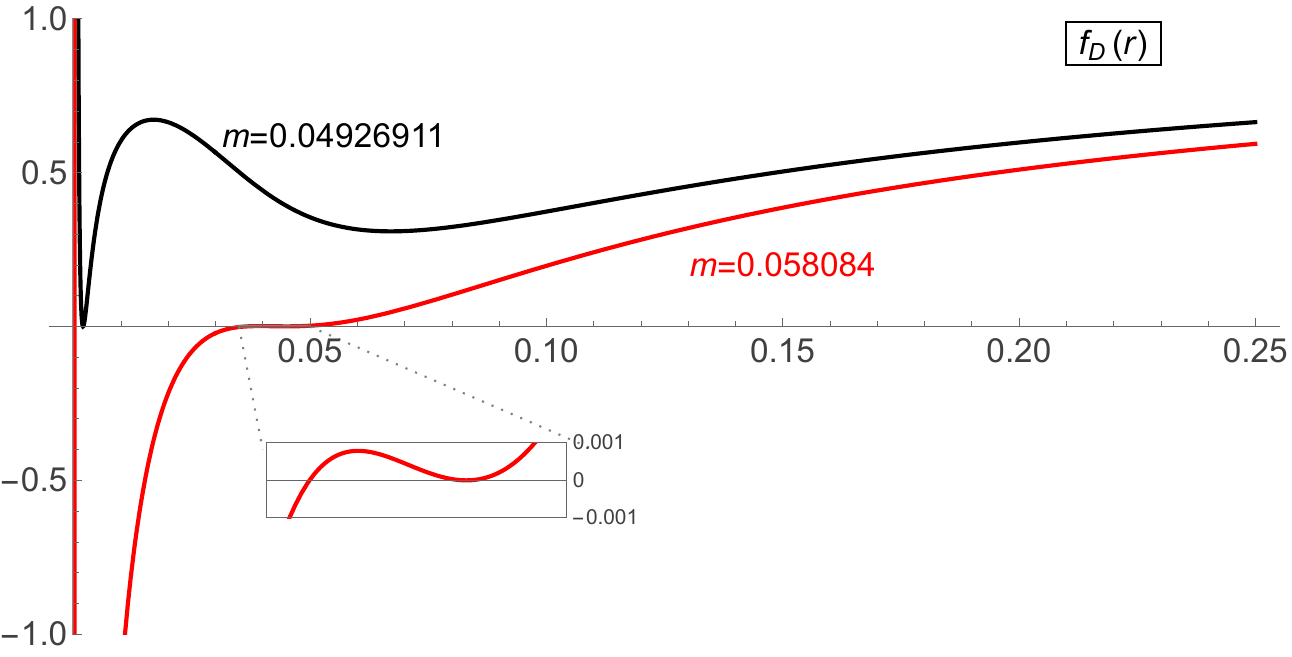}
\caption{A plot of $f_D(r)$ for the dyonic topological solution in the parameter region where the lapse function has turning points outside the horizon. Here, we show the two extremal solutions for the same charges $p = 0.002, q = 0.06, b = 0.1$ with masses as labelled. By varying $m$ the discontinuous horizon behaviour is observed.}
\label{fig:Qextremals}
\end{figure}

Unlike the magnetic black hole above, this solution requires the mass to be chosen carefully in order to have a horizon, as the $q^2$ term is always positive and divergent at the origin (along with $p^2/r^2)$. We therefore have at least one extremal limit associated to each pair of $p,q$, and for small mass and charge there is the possibility of two distinct extremal solutions as indicated in figure \ref{fig:Qextremals}.

\section{Geodesics}
\label{sec:geodesics}

Having discussed the solutions with non-monotonic behaviour of the lapse function, we now turn to the geodesics in the geometry. 
We anticipate that the presence of the turning points will lead to interesting orbits around the black holes due to the weighting of angular momentum in the geodesic equations, and that there may be additional light-rings outside the horizon. 
However, it turns out that the structure of light orbits in particular becomes extremely rich due to the nonlinear interactions of the photon that can be encoded in effective optical metrics \cite{Obukhov:2002xa, Novello:1999pg, Ovgun:2025mdg, DeLorenci:2000yh} that produce significantly different results for the two polarisation states in the vicinity of the horizon.

First, recall that without loss of generality one can take geodesics restricted to the equatorial plane, deriving the standard geodesic equation of motion as
\begin{equation}
\frac{1}{2}\dot{r}^2+V_\mathrm{eff}(r)=\frac{E^2}{2},
\end{equation}
where $V_{\mathrm{eff}}$ is the effective potential, 
\begin{equation}
V_{\mathrm{eff}}=\left(\frac{L^2}{r^2}+\eta\right)\frac{f(r)}{2},
\label{effectivepotential}
\end{equation}
$L = r^2 \dot{\phi}$ represents the angular momentum, and $E = f \dot{t}$ the energy of the particle.
The variable $\eta$ encodes whether the geodesic is timelike ($\eta=1$) or null ($\eta=0$).

\subsection{Null geodesics}

Null trajectories correspond to $\eta=0$, thus the effective potential takes the particularly simple form
$V = L^2 f(r) / (2r^2)$. We can always rescale the affine parameter of a null geodesic, thus we could, without loss of generality, choose to set $L$ or $E$ equal to $1$.  
Now consider the magnetic geometry described by the analytic expression \eqref{eqn:f(r)}. The effective potential is then
\begin{equation}
V_{\text{null}}(r) =\frac{L^2}{2}\left(\frac{1}{r^2}-\frac{2m}{r^3}+\frac{p^2}{r^4}-\frac{2ap^4}{5r^8}\right)
= V_{\text{RN}} 
- \frac{L^2 a p^4}{5 r^8} \,,
\label{nulleffpotl}
\end{equation}
where $V_{\text{RN}}$ is the form of the effective potential for the RN spacetime.

Taking $a$ to be positive, this shows that in the NLED geometry, there are a wider range of orbits for 
a given energy. We now show that if the metric potential $f(r)$ has turning points outside the horizon, there exist additional turning points in $V_{\text{null}}$, hence additional null rings with one being stable. 

\begin{figure}
\centering
    \includegraphics[width=0.8\textwidth]{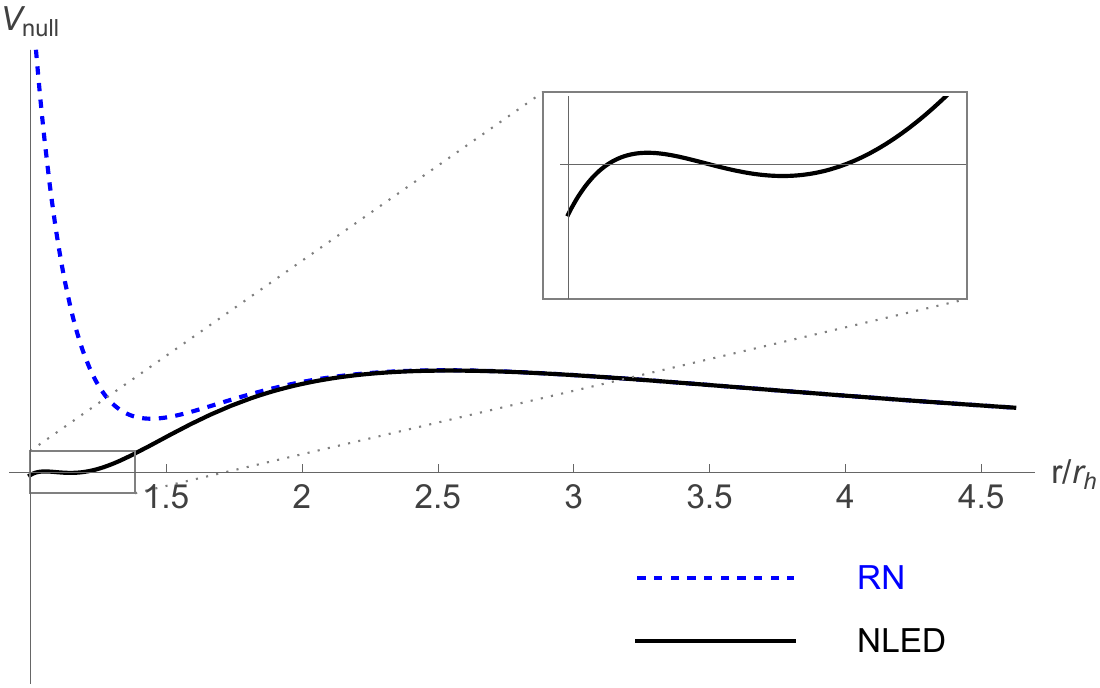}
    \label{fig:V(r)RNnull}
\caption{Comparison of effective potentials for the geometric null geodesics in the NLED geometry (in black) with the Reissner-Nordstr\"om potential (blue dashed). For both the RN and NLED geometrical case, varying $E$ simply changes the zero-point of the effective potential, so the existence of two additional null rings, one stable and one unstable, can be easily seen. The parameters used in this plot are: $m = 1.1464, p = 1.17, a = 0.1, E^2/L^2 = 0.0015$.}
\label{fig:RNvNLED}
\end{figure}

As with the RN geometry, $V_{\text{null}}$ vanishes at the horizon and infinity, with $V_{\text{null}}$ being positive inbetween. Thus, there is always at least one turning point of $V_{\text{null}}$ outside the horizon that will be a local maximum, giving rise to an unstable null orbit. 
For the RN spacetime, this is easily seen to be at the radius\footnote{It is worth noting that the expression for the RN null orbit radius \eqref{RNnullr} is well defined for $m<p<3m/2\sqrt{2}$, even though the RN geometry loses
its horizon for $p>m$. }
\be
r_{\text{RN}} = \frac12 \left [ 3m + \sqrt{9m^2 - 8p^2} \right ].
\label{RNnullr}
\ee
With the modified potential \eqref{nulleffpotl}, we see that the canonical RN null ring is actually perturbed outwards, as
\be
V_{\text{null}} ' = V_{\text{RN}}' + \frac{8aL^2 p^4}{5r^9},
\ee
thus $V_{\text{null}} '$ is slightly positive at the location of the RN null ring. For small $a$ this shift is straightforwardly derived as
\be
\delta r_{\text{no}} = \frac{8p^4\,a}{5 r_{\text{RN}}^3(r_{\text{RN}}^2 - 2p^2)} + {\cal O}(a^2)
\ee
For the values of $p$, $m$ and $a$ we consider, this correction remains small, and we conclude that there remains an unstable null orbit in the NLED spacetime close to the RN value (see figure \ref{fig:RNvNLED}).

Now suppose we are in a parameter regime with $f$ undulating near the horizon, i.e.\ $f'\leq0$ for a range of $r$ with $f>0$ in that range. 
Clearly, while $f'\leq0$ with $f>0$, $V'_{\text{null}}<0$, hence $V_{\text{null}}$ must have a turning point between the horizon and this first turning point of $f$ as $V$ increases away from the origin. Since we have already shown that $V$ has the standard ``null ring'' maximum further away from the horizon than $r_{RN}$, this implies that there must also exist a local minimum of $V_{\text{null}}$. 
In figure \ref{fig:RNvNLED} we show a comparison between the effective potential in RN and NLED for the analytic magnetically charged metric (left plot).  

This discussion so far has simply taken the geometric structure of the magnetic black hole and considered the null geodesics therein. However, the usual null trajectories one is interested in are those of photons, and the propagation of light is impacted by the nonlinearity of the electromagnetic Lagrangian \cite{Obukhov:2002xa, Novello:1999pg, Ovgun:2025mdg, DeLorenci:2000yh}.

Following the method of characteristics (see Appendix \ref{app:characteristics}), and noting that the mixed derivative $\partial_{FG}$ of the electromagnetic Lagrangian, ${\cal L} = F - aF^2 - b G^2$, vanishes, we find that the optical effective metrics are:
\be
\texttt{g}^{\mu\nu}_{\pm} = \left ( {\cal L}_F + \frac{-\Omega \pm \sqrt{\Delta}}{2{\cal L}_{GG}} \right ) g^{\mu\nu} + 4{\cal L}_{FF} F^\mu_{~\sigma} F^{\nu\sigma} 
\ee
where 
\be
\beal
\Omega &= {\cal L}_F ( {\cal L}_{GG} - {\cal L}_{FF} )
+ 2 F {\cal L}_{FF} {\cal L}_{GG}\\
\Delta &= \Omega^2 + 4 G^2 {\cal L}_{GG}^2 {\cal L}_{FF}^2
\eeal
\ee

For the magnetic black hole background, $G=0$ \footnote{Note that even though $G=0$ in the background, $G$ will not vanish for a general electromagnetic perturbation.}, and these metrics reduce to:
\be
\beal
\texttt{g}_{+}^{\mu\nu} &= (1-2aF) g^{\mu\nu} - 8a F^\mu_{~\sigma} F^{\nu\sigma} \\
\texttt{g}_{-}^{\mu\nu} &= (1-2aF - 4bF) g^{\mu\nu} - 8b F^\mu_{~\sigma} F^{\nu\sigma}
\eeal
\label{opticalmetrics}
\ee
where $F = 2 p^2/r^4$ for the magnetic black hole.

\begin{figure}
\centering
\includegraphics[width=0.8\textwidth]{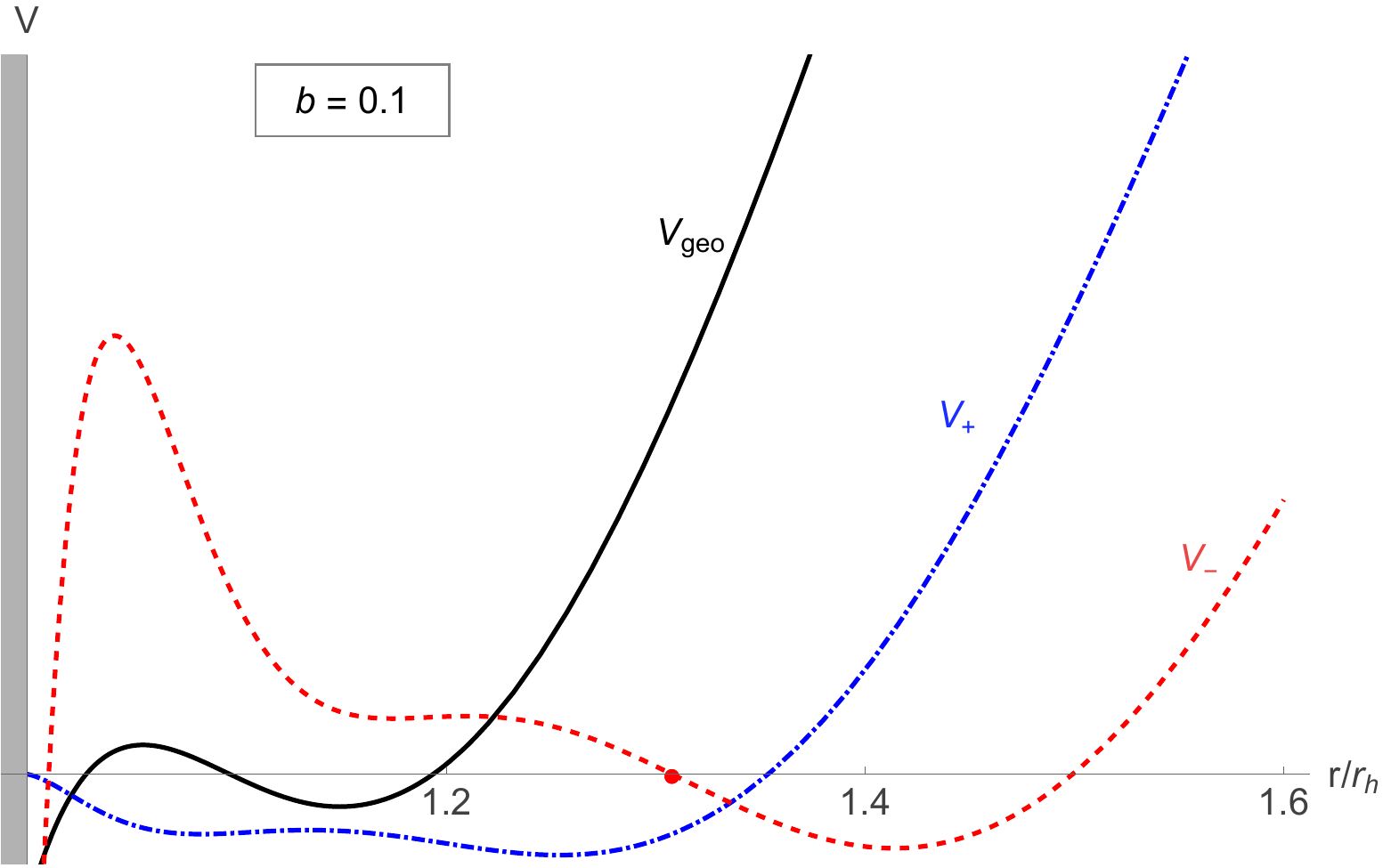}
\caption{A comparison of the optical potential for the NLED spacetime versus the pure geometric potential for the test case $b= 0.1$ (see figure \ref{fig:varyb} for the impact of varying $b$). The red dot indicates the threshold radius. The outer event horizon is indicated by a thick grey line on the left, and the radius is shown in units of $r_h$. The parameters used in this plot are: $m = 1.1464, p = 1.17, a = 0.1, E^2/L^2 = 0.0015$.}
\label{fig:Vcomparisons}
\end{figure}

\begin{figure}
\begin{subfigure}{0.48\textwidth}
    \includegraphics[width=\textwidth]{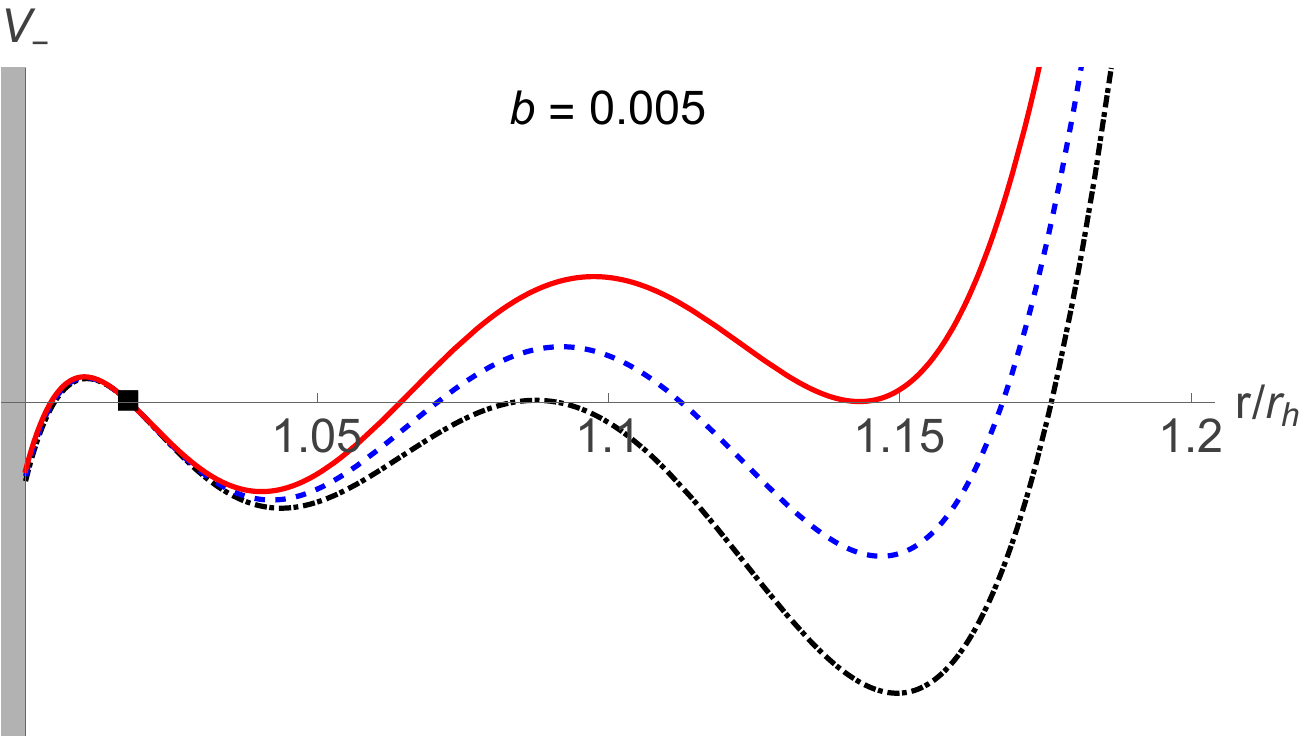}
\end{subfigure}
\hfill
\begin{subfigure}{0.48\textwidth}
    \includegraphics[width=\textwidth]{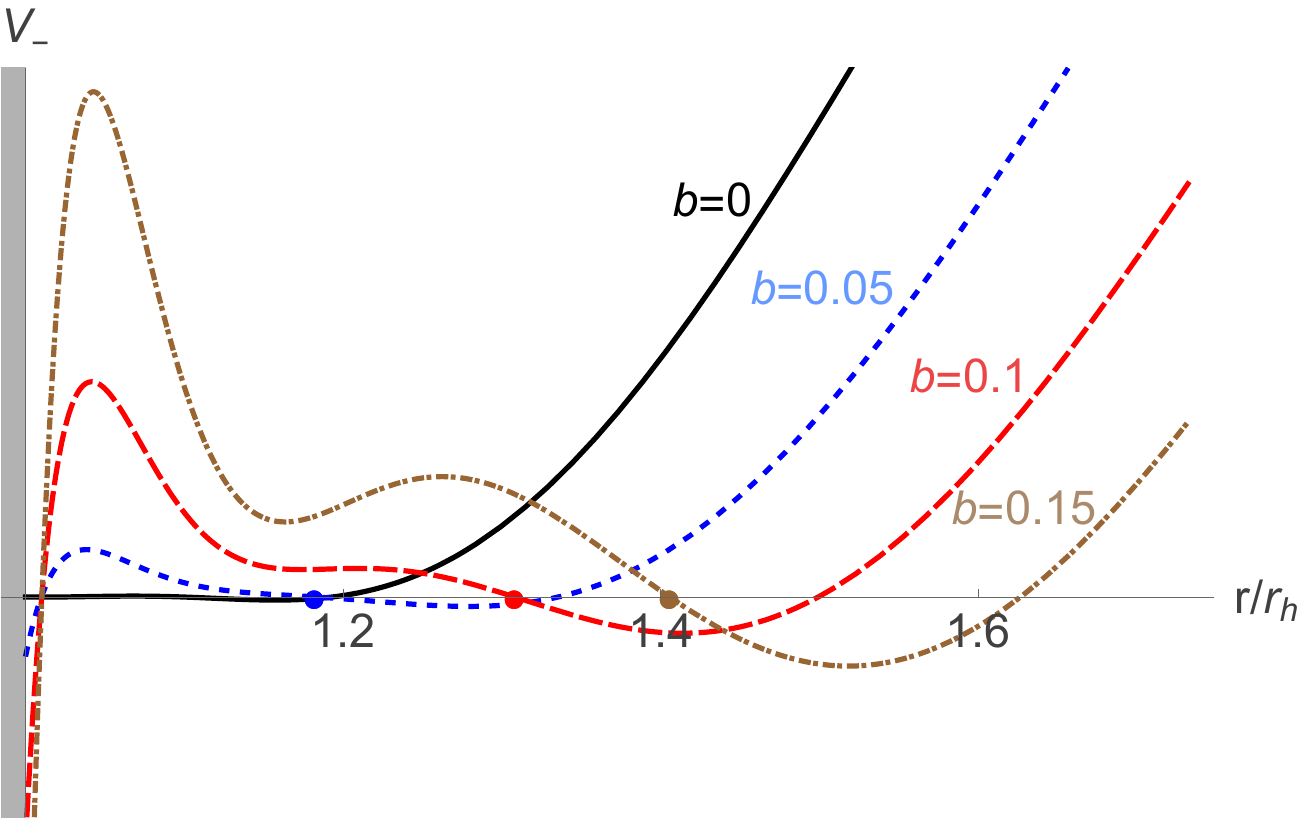}
\end{subfigure}
\caption{Plots of the $\texttt{V}_-$ effective potential illustrating the range of light trajectories. For all $b$, there are trapped trajectories near the horizon, with the specific range of radius given by $\texttt{E}_-/\texttt{L}_-$.
{\bf Left:} Illustration of multiple trapped regions and stable light-rings at small $b=0.005$.  As $\texttt{E}_-/\texttt{L}_-$ increases, $\texttt{V}_-$ becomes more negative, displaying different possible band behaviour. Here, the range of $(\texttt{E}_-/\texttt{L}_-)^2$ is $0.0112$ (red), 0.00105 (blue-dashed), and $0.000968$ (black dot-dashed). The threshold radius $a+2b = 1/2$ is indicated by a black square, and the outer event horizon by the thick grey line.
{\bf Right:} Plot illustrates how varying $b$ impacts the optical potential $\texttt{V}_-$ at fixed $\texttt{E}_-^2/\texttt{L}_-^2 = 0.0015$. The threshold radius is indicated for each labelled curve.
(The parameters used in both plots are: $m = 1.1464, p = 1.17, a = 0.1$.)}
\label{fig:varyb}
\end{figure}

Returning to the optical light ray equations, these metrics cause a renormalisation of the energy and angular momentum parameters: 
\be
\beal
E_+ &= \frac{f \dot{t}}{(1 - 2aF)} \quad &L_+ &= \frac{r^2 \dot{\phi}}{(1 - 6aF)} \\
E_- &= \frac{f \dot{t}}{(1 - 2aF-4bF)} \quad &L_- &= \frac{r^2 \dot{\phi}}{(1 - 2aF - 8bF)} 
\eeal
\ee
thus for the same (affine) rate of change of $t$ and $\phi$, we have different effective energies and angular momenta.
These in turn modify the effective potentials, however, there is a key difference in how the energy parameter feeds into the potential: For $V_{\text{null}}$, $E^2$ is simply a constant that sets the zero of the effective potential, and the type of allowed orbits can be deduced by shifting the axis up or down. Here however, the {\emph{shape}} of the potential is impacted by the value of $E_\pm$, particularly near the horizon (see figure \ref{fig:Vcomparisons} and figure \ref{fig:varyb} for example), therefore $E$ does not simply set a zero level for $\dot{r}$. We therefore include $E^2$ in the effective potentials so that 
\be
\frac{\dot{r}^2}{2} + \texttt{V}_\pm = 0
\ee
and $\texttt{V}_\pm=0$ now corresponds to $\dot{r}=0$. With this re-zeroing, the effective potentials are found to be:
\be
\beal
\texttt{V}_+ &= - \frac{E^2_+}{2}( 1 - 2aF)^2 + \frac{L^2_+ f}{2r^2} (1-2aF)(1-6aF) \\
\texttt{V}_- &= - \frac{E^2_-}{2}\left ( 1 - 2(a+2b)F \right )^2 + \frac{L^2_- f}{2r^2} (1-2(a+2b)F)(1-2(a+4b)F)
\eeal
\ee
where we are now including the constant $E_\pm^2$ in the effective potential.

In each case, the presence of these extra factors involving $F$ acts to alter the relation of the energy and angular momentum constants relative to the variation of the time and angular coordinates which then feeds in to a different weighting of the contribution of each of these in the effective potential. 
Since $F = 2p^2/r^4$, these extra terms only become relevant close to the horizon, however, in this region they can have a significant effect.
In the geometric potential, $E$ always contributes negatively and $L$ positively, however, if $1>2aF>1/3$, then {\it both} the energy and angular terms in $\texttt{V}_+$ become negative.
 
To determine the behaviour of these factors, we refer to the discussion in the appendix of \cite{Croney:2025}, where bounds on the value of $p$ (and, implicitly, $m$) were obtained. 
The dominant energy condition implies that $(1-2aF)>0$ throughout the spacetime, however for the allowed range of $p$, the upper limit of $m$ for nonmonotonicity of $f$ implies that $(1-6aF)<0$, hence
$\texttt{V}_+$ is strictly negative immediately outside the horizon for the case of oscillating $f$, thus there are no bound orbits for this polarisation, however the unstable null orbit in the vicinity of the RN null radius remains.

\begin{figure}
\centering
\includegraphics[width=0.8\textwidth]{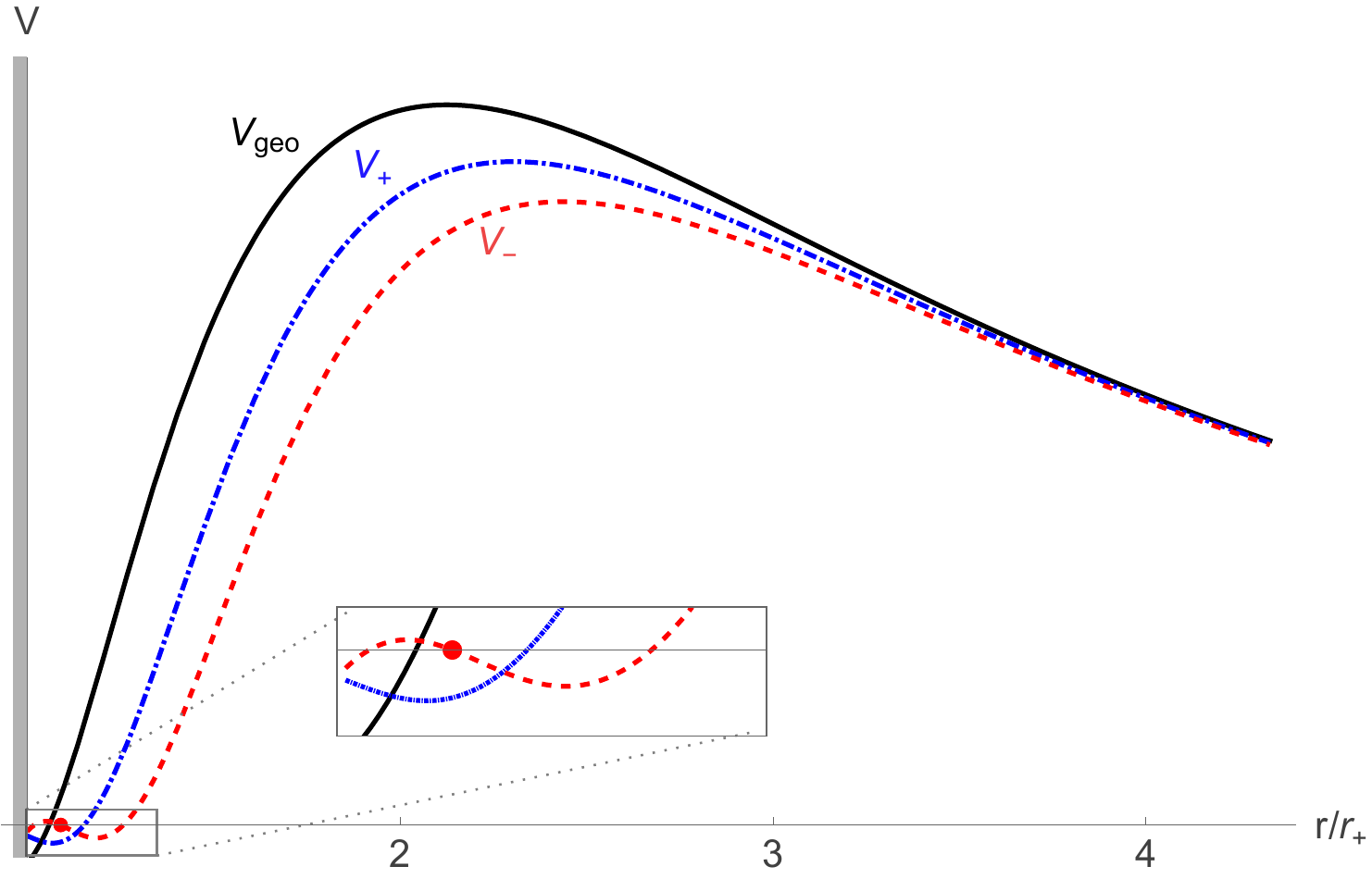}
\caption{The effective potentials for $m>m_{\text{crit}}$, where $f$ is now monotonic. While the geometric and $V_+$ potentials now show no trapped geodesics, the second optical potential still has a small region of trapped null orbits very near the horizon. The outer event horizon is indicated by the thick grey line.
The parameters used in this plot are: $m = 1.1473, p = 1.17, a = b = 0.1, E/L = 0.05$.}
\label{fig:mplus}
\end{figure}

For $\texttt{V}_-$, there is more complex behaviour, depending on the relative values of $\texttt{E}_-$ and $\texttt{L}_-$, as well as on $b$. 
If $b=0$, $\texttt{g}_-$ is conformal to the geometric metric, which means that the limiting radii of any bound orbits are identical, however as the shape and magnitude of the potential are impacted by these multiplicative factors, the details of these trapped geodesics differ.
For nonzero $b$, if $(1-2(a+2b)F)=0$ at some $r_b$, then due to the definition of $\texttt{E}_-$ it is easy to see that $r_b$ will be an inner turning point for any geodesics, which therefore will be repelled from the horizon. 
Further, this feature will be present for all $b> r^4/8p^2 -a/2$, therefore there are always trapped light rays outside the horizon as well as the usual unstable light-ring, corresponding to various values of $\texttt{E}_-/\texttt{L}_-$, see figure \ref{fig:varyb}.
In fact, due to this critical threshold, for sufficiently large $b$ ($\sim 0.05$ for our choice of $p,a$) there can be trapped light orbits outside the horizon even if $f$ is monotonic, as illustrated in figure \ref{fig:mplus}.

To sum up, the geometry has, in addition to the standard black hole null ring near the RN value, regions of trapped null geodesics as well as two additional inner null rings, one of which is stable. Light however reacts to the nonlinearity of the electromagnetic interaction and follows distinct trajectories determined by the geodesics in the polarisation optical metrics \eqref{opticalmetrics}. One polarisation has a single unstable light-ring, however the other polarisation state has a rich structure, with multiple trapped geodesic bands, stable and unstable light-rings. 

For the quasi-topological metric \eqref{dyonicf}, a similar behaviour is observed for the geometric null geodesics, with an unstable null ring close to an expected RN value plus two other circular null orbits close to the horizon. 
Computing the optical metrics following \cite{Novello:1999pg} gives
\be
\beal
\texttt{g}_+^{\mu\nu} &= g^{\mu\nu} \\
\texttt{g}_-^{\mu\nu} &= g^{\mu\nu}
-8b F^{\mu}_{~\sigma} F^{\nu\sigma} + 4b F g^{\mu\nu}
\eeal
\ee
leading to the effective potentials:
\be
\beal
\texttt{V}_{D+} &= - \frac{E^2_+}{2} + \frac{L^2_+ f}{2r^2} \\
\texttt{V}_{D-} &= - \frac{E^2_-}{2}\left ( 1 + \frac{8 b p^2}{r^4} \right )^2 + \frac{L^2_- f}{2r^2} \left ( 1 + \frac{8 b p^2}{r^4} \right )\left ( 1 - \frac{8 b q^2 r^4}{(r^4+8 b p^2)^2} \right )
\eeal
\ee
While the details of the geodesics vary, the qualitative aspects of multiple null orbits for the plus potential, and trapped geodesics near the horizon for the minus potential remain.


\subsection{Timelike geodesics}

For the sake of completeness, we briefly note the behaviour of timelike geodesics. An immediate observation is the existence of stable, static observers outside the horizon when there are additional turning points in $f$: With $L=0$, $\dot{r}^2 = E^2 - f$, thus picking $r_s$ to be the location of the middle (local minimum) turning point of $f$ outside the horizon and fixing $E = \sqrt{f(r_s)}$ gives a stable timelike geodesic $r = r_s$.

In \cite{Wang:2025wob}, an analysis of orbiting timelike geodesics for a similar lapse function that has turning points outside the horizon is presented. While we have been unable to reproduce their particular geometry for their stated nonlinear interaction term, the qualitative features of the timelike geodesics agree.

\begin{figure}
\centering
\includegraphics[width=0.8\textwidth]{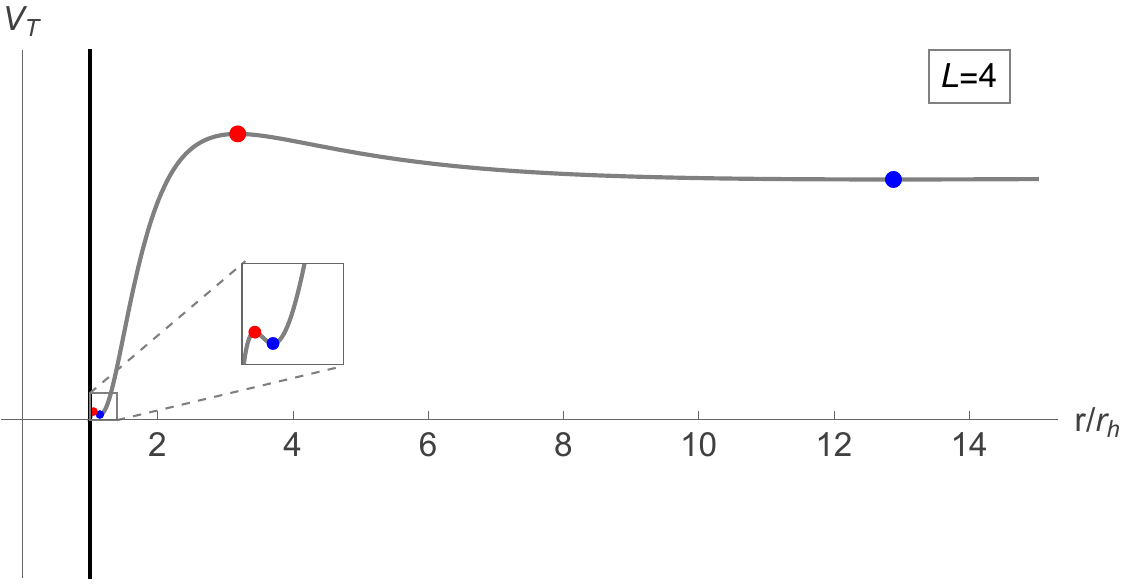}
\caption{Plot of the effective potential $V_T$ for timelike geodesics for a representative value of the angular momentum constant $L=4$, where there are four stationary points outside the horizon, corresponding to two unstable timelike orbits and two stable timelike orbits. The solid vertical line marks the location of the event horizon, and the dots mark the stationary points outside the horizon. The red dots represent unstable circular orbits and the blue dots stable ones. The parameter values are $m = 1.1464, p = 1.17, a = 0.1$.}
\label{fig:Vtimelike}
\end{figure}

For timelike geodesics in our magnetic geometry, we set $\eta=1$ in \eqref{effectivepotential} to obtain
\begin{equation}
V_{T}(r)=\left ( 1 + \frac{L^2}{r^2} \right ) \frac{f(r)}{2} = \frac{1}{2}-\frac{m}{r}+\frac{L^{2}+p^{2}}{2r^{2}}-\frac{L^{2}m}{r^{3}}+\frac{L^{2}p^{2}}{2r^{4}}-\frac{ap^{4}}{5r^{6}}-\frac{aL^{2}p^{4}}{5r^{8}}
\end{equation}
For the general case with angular momentum $L$, note that since $V_T$ is zero with positive derivative at the horizon, and tends to $1/2$ from below at infinity, turning points in $V_T$ must occur in pairs. 
As with the RN black hole there exists a threshold $L_{\mathrm{ISCO}}$ above which there are two circular orbits: One inner unstable orbit, and an outer stable orbit close to the RN values. However, due to the turning points in $f$, irrespective of $L$ there are also always a pair of circular orbits (again inner unstable and outer stable) close to the horizon. 
To see this, note that $V_T$ is zero with positive derivative at the horizon, however $V_T$ must have a turning point prior to the first turning point of $f$ since $V_T' = (1+L^2/r^2)f'-2L^2 f/r^3<0$ here. Since turning points occur in pairs, we must therefore have a corresponding stable circular orbit close to the horizon (see figure \ref{fig:Vtimelike}).
As is the case with Schwarzschild, however, timelike orbits are found to lie outside the unstable (geometric) null ring.

\section{Quasinormal modes}\label{sec:qnms}


We now turn to the QNM spectrum of the dyonic black hole solutions, which characterises the linear response of the spacetime to perturbations and governs the ringdown. Given the rich geodesic structure of the spacetime, we also expect the QNM spectrum to reflect the existence of additional turning points in the lapse function. We focus on scalar perturbations, starting with introducing the setup and numerical method, before presenting our results. We note that the geodesic analysis of the previous section and the QNM analysis here probe different aspects of the spacetime. While the optical metrics govern the propagation of electromagnetic waves, the quasinormal modes considered here correspond to a scalar field perturbation of the background geometry and would therefore be more naturally associated to the geometric geodesic analysis. Although these sectors are distinct, the richness of the phenomena we observe share a common origin, which is the non-monotonicity of the lapse function. Therefore, we expect the results presented here to extend to vector and tensor perturbations.

First, we consider a test field perturbation of the fixed background spacetime. In particular, we look at a massless scalar field $\Psi(t,\mathbf{r})$, governed by the Klein-Gordon equation,
\begin{equation}
    \frac{1}{\sqrt{-g}}\partial_\mu\left(\sqrt{-g}\, \partial^\mu \Psi \right)=0.
\end{equation}

Given the time-independence and spherical symmetry of the metric, the scalar field has a natural decomposition,
\begin{equation}
    \Psi(t,r,\theta,\phi) = \sum_{lm_\phi} e^{-i\omega t}R(r)Y_{lm_\phi}(\theta,\phi),
\end{equation}
where $Y_{lm_\phi}(\theta,\phi)$ are the spherical harmonics, and $l$ and $m_\phi$ are the polar and azimuthal indices. This leads to a radial equation of the form,
\begin{equation}\label{eqn:radial}
    \left[\frac{d^2}{dr_*^2}+\left(\omega^2-V_l(r)\right)\right]R(r)=0,
\end{equation}
where $V_l(r)$ is the effective potential given by,
\begin{equation}\label{eqn:V_l(r)}
    V_l(r)=f(r) \left[\frac{l(l+1)}{r^2}+\frac{f'(r)}{r}\right],
\end{equation}
with $r_*$ being the ``tortoise'' coordinate defined by,
\begin{equation}
    \frac{dr_*}{dr} = \frac{1}{f(r)}.
\end{equation}

QNMs are complex frequency solutions of \eqref{eqn:radial} that satisfy ingoing boundary conditions at the black hole event horizon, and outgoing boundary conditions at spatial infinity,
\begin{equation}
\begin{split}
    &R(r) \sim e^{-i\omega r_*},\quad r_*\rightarrow -\infty, \\
    &R(r) \sim e^{+i\omega r_*},\quad r_*\rightarrow +\infty.
\end{split}
\end{equation}

\subsection{Numerical method}
To solve the radial equation, we employ a pseudospectral collocation method on a compactified domain (see, e.g. \cite{Boyd:2001} for a general overview, and \cite{Jansen:2017} for applications to QNMs).
We work with the compactified coordinate $z\in[0,z_h]$, defined by $z:=1/r$, with $z_h:=1/r_h$. With this choice of coordinate, the radial equation \eqref{eqn:radial} transforms to,
\begin{equation}\label{eqn:radialz}
    z^4 f(z)^2 \frac{d^2R(z)}{dz^2}+z^3 f(z)\left[2f(z)+zf'(z)\right] \frac{dR(z)}{dz} + \left[\omega^2+z^3 f(z) f'(z)\right] R(z)=0,
\end{equation}
where the metric function has been appropriately transformed, 
\begin{equation}
    f(z)=1-2mz+p^2z^2-\frac{2}{5}ap^4z^6,
\end{equation}
and for concreteness we have fixed $l=0$, but we have also verified that the qualitative features of the effective potential remain the same for nonzero $l$.

We scale out the asymptotic behaviour of $R(z)$,
\begin{equation}
    R(z)=A(z)B(z)\psi(z),    
\end{equation}
where the prefactors $A(z)$ and $B(z)$ encode the behaviour of the scalar field near the horizon ($z=z_h$), and near spatial infinity ($z=0$), respectively,
\begin{equation}
    \begin{split}
        A(z) &= e^{i\frac{\omega}{z}} z^{-i(2m\omega)-1},\\
        B(z) &=\left(1-\frac{z}{z_h}\right)^{-i\frac{\omega}{f'(r_h)}-1}.
    \end{split}
\end{equation}

Note that the extra powers of $-1$ in $A(z)$ and $B(z)$ are to ensure that $\psi(z)$ decays linearly at each boundary, which helps with the implementation of boundary conditions and numerical stability, following \cite{Jansen:2017}.

To solve for the quasinormal mode spectrum, we employ the pseudospectral method with 300 Chebyshev-Gauss-Lobatto (CGL) nodes and a numerical working precision of 300. The differential equation \eqref{eqn:radialz} is discretised using Chebyshev differentiation matrices, and recast as a generalised eigenvalue problem,
\begin{equation}
\mathbf{A}\vec{v}=\omega\mathbf{B}\vec{v}.
\end{equation}
We solve this directly using Mathematica's built-in Eigenvalues function. This inevitably produces some spurious modes, mixed in with the actual quasinormal modes, however we can distinguish between them by solving our problem with different numbers of CGL nodes as follows. We repeat the calculations with 220, 240, 260, and 280 CGL nodes, and while the spurious modes move (on the complex plane) regularly and predictably with our number of nodes, the quasinormal mode frequencies converge to the true value as we increase the number of nodes.

\subsection{Results}
\begin{figure}
\begin{subfigure}{0.48\textwidth}
    \includegraphics[width=\textwidth]{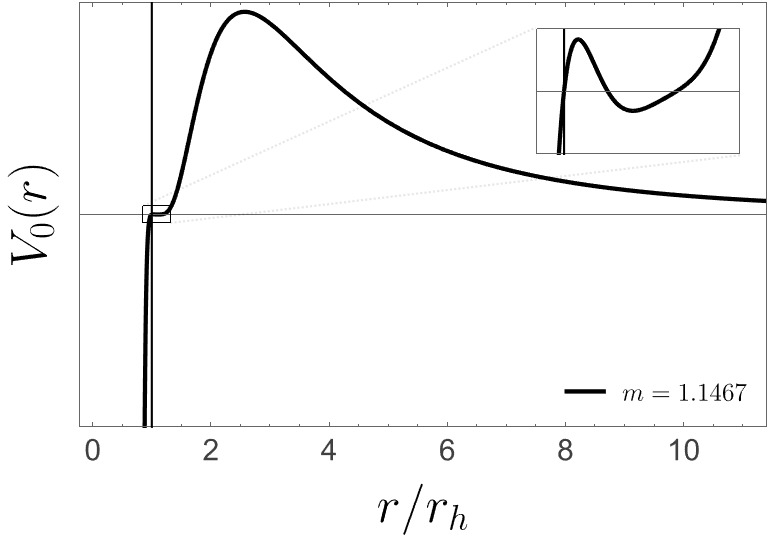}
    \label{fig:qnmV(r)2branch}
\end{subfigure}
\hfill
\begin{subfigure}{0.48\textwidth}
    \includegraphics[width=\textwidth]{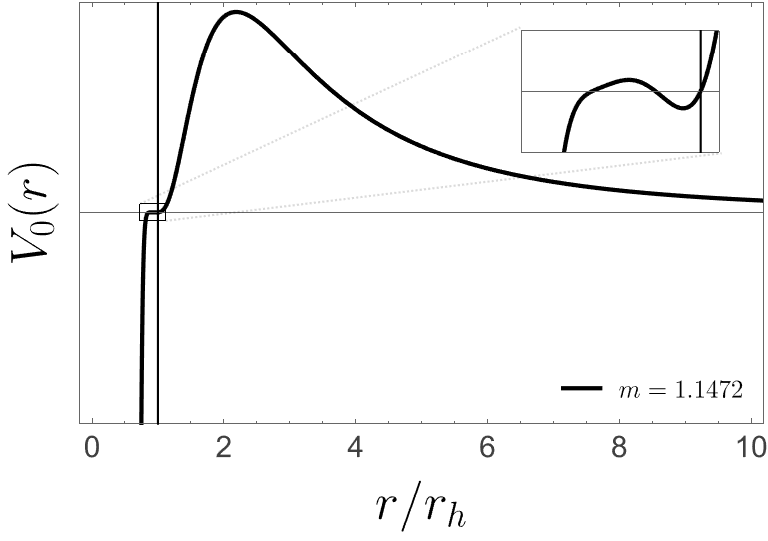}
    \label{fig:qnmV(r)1branch}
\end{subfigure}
\caption{Plots of the effective potential in the wave equation governing the radial part of the scalar field, for $q=0, p=1.17, a=0.1$. The vertical line marks the location of the black hole event horizon. {\bf Left:} the local minimum in the potential lies outside the horizon, and so it is ``seen'' by the scalar field. {\bf Right:} the increased mass has shifted the horizon, thus the minimum in the potential is ``hidden'' behind it.}
\label{fig:qnmV(r)}
\end{figure}

Typically, one obtains a single branch of QNMs for a (sub-extremal) black hole in the Kerr-Newman family. However, this is modified in our setup in the regions of parameter space where the lapse function $f(r)$ contains a turning point outside the black hole horizon. This additional turning point is directly responsible for the appearance of a minimum in the effective potential $V_l(r)$, as shown in figure \ref{fig:qnmV(r)}, acting as a trapping region for perturbations. We find a secondary, distinct branch of QNMs due to resonances associated with this feature.

\begin{figure}
    \centering
    \includegraphics[width=0.8\textwidth]{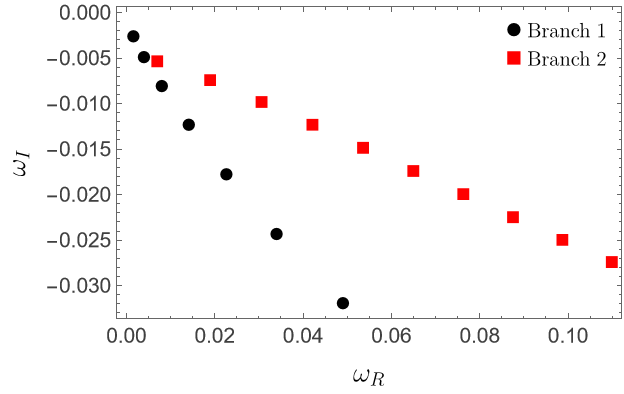}
    \caption{A plot showing QNM frequencies for $q=0,p=1.17,a=0.1,m=1.1467$. There are two distinct branches of modes. The modes on branch 2 are in general longer-lived than those on the first branch (for a given overtone number), as they have a smaller $|\omega_I|$, and hence, decay more slowly.}
    \label{fig:qnmFreq2branch}
\end{figure}

In figure \ref{fig:qnmFreq2branch} we show the mode frequencies for both branches at a fixed mass $m$. This new branch of QNMs is longer lived (with smaller $|\omega_I|$) than the usual branch, as is expected for trapped modes. As we vary the parameters of our system, the minimum in the effective potential disappears behind the horizon and the secondary branch vanishes, verifying that this branch is indeed associated with the non-monotonicity of $f(r)$.

Although these results were found for purely magnetic black holes, we expect similar branching behaviour in the dyonic case with small electric charge, but not the pure electric case. These cases cannot be solved for analytically in NLED in full generality, however we have scanned the parameter space numerically and found the following features.

Qualitatively, purely electric NLED black holes $(p=0)$ resemble the RN case, but allow the electric charge $q$ to exceed the mass $m$ while still retaining a horizon. However, unlike the purely magnetic case, where a horizon always exists, there remains an extremal limit: beyond a critical charge (typically only a few percent above $m$ in Planck units), the horizon disappears and a naked singularity forms. Increasing the non-linear coupling $a$ extends the range of $q>m$ for which a horizon persists, though a cut-off always remains. In this electric case, the “jumping horizon” phenomenon appears absent, suggesting a single branch of QNMs. By contrast, in the dyonic case, the jumping horizon behaviour persists, and increasing $q$, $a$, or $b$ enhances deviations from RN. Here, both $q$ and $p$ can be increased without bound while maintaining a horizon, although large regions of parameter space violate the DEC when these parameters grow simultaneously. Accordingly, we expect the qualitative features of our analysis to carry over, including the presence of a secondary QNM branch. 

\section{Discussion}
\label{sec:discussion}

In this paper, we have explored the consequences of the exotic 
geometries that can arise in NLED black holes. Taking the analytic 
magnetic black hole solution \cite{Croney:2025} as an exemplar, we 
have found near horizon additional null rings and near-horizon 
static observers. We also noted the rich structure of light rays 
near the horizon using the effective optical geometry approach. 
While this approach is a useful shorthand for deriving the 
main behaviour of photon trajectories via the characteristics of
the electromagnetic perturbation equations, it should be noted that
if the geodesics become too strongly curved the applicability of 
this method is not so clear \cite{Schellstede:2016zue}.

The additional structure of the geodesic effective potential is hidden behind the outer barrier associated with the unstable light-ring, and therefore may not be accessible to generic trajectories. Nevertheless, partial transmission through the barrier, or scatter of photons from the local environment, could allow this region to become populated over large timescales. This raises the possibility of an instability as photons become trapped in this region, leading to an accumulation of energy density, as suggested in \cite{Cardoso2014,Cunha_2017, cunha2025}. 
We explored the impact of a shell of radiation around the black hole, which is readily computed using the Israel junction conditions \cite{Israel:1966rt}. The radiation wall, with an equation of state $p = \rho/2$, matches an inner magnetic geometry to an outer one with mass incremented by the energy in the shell. This indicates that as the shell accumulates energy, the outer mass increases until there is a critical mass at which the horizon of the exterior geometry can no longer sit inside the shell as the ``jumping horizon'' phenomenon has taken place. Of course, the horizon does not literally ``jump'', rather, the teleological nature of the black hole horizon means that the shell is consumed by the expanding black hole from the point of view of an asymptotic observer. We leave the calculation of instability timescales to future work.

Another possible instability could arise from the near-horizon
static timelike observers. Similar to the shell of radiation 
discussed above, a static shell of dust can also exist outside
the horizon, matching two NLED spacetimes with differing masses. 
As the density of this shell is increased, the same ``jumping
horizon'' behaviour is observed, although it is perhaps less 
credible that matter might be scattered in such a way as to get 
trapped so close to the horizon.

The double-barrier structure of the geodesic effective potentials arising from the extrema in the lapse function places our system within a broader class of models exhibiting similar features, including certain scalar-tensor theories \cite{vlachos2021}, black holes in massive gravity \cite{Dong:2021}, a number of exotic compact object models \cite{cardoso2016,bueno2018,Churilova2021}, and more recently in black holes endowed with additional primary hair \cite{Konoplya:2026}. In such cases, the potential contains a cavity in which perturbations can become temporarily trapped before gradually leaking out, giving rise to long-lived resonances and a sequence of late-time echoes in the ringdown signal. Given that our setup also contains a trapping region between unstable null rings, we naturally expect qualitatively similar phenomenology.

Finally, we remark on the spectral instability of QNMs caused by a small perturbation to the effective potential \cite{Nollert:1996rf}. As we have shown, a relatively miniscule additional feature in the effective potential governing QNMs causes a dramatic change in the QNM spectrum, reinforcing the findings in the literature \cite{Jaramillo:2020tuu,Cheung:2021bol}. While the spectra are destabilised, this does not necessarily translate into a physical instability of the time-domain signal \cite{Berti:2022xfj}. A detailed investigation would require time-domain simulations, which we leave for future work.

\begin{acknowledgments}
We are grateful to Jorge Russo and Paul Townsend for useful remarks.
LC and AG were supported by King's College London through an NMES funded studentship. RG is supported in part by the STFC grant (ST/X000753/1) and in part by the Perimeter Institute for Theoretical Physics. CRV acknowledges support from the Secretar\'ia de Educaci\'on, Ciencia, Tecnolog\'ia e Innovaci\'on de la Ciudad de M\'exico (SECTEI) of Mexico City. Research at Perimeter Institute is supported in part by the Government of Canada through the Department of Innovation, Science and Economic Development and by the Province of Ontario through the Ministry of Colleges and Universities.
\end{acknowledgments}

\appendix

\section{Classification of horizons}

Here we discuss the behaviour of the lapse function for the magnetic black hole and give a physical classification of each of the horizons when more than one are present.
We use the Sturm method \cite{Basu2006} to determine the number of roots of $f(r)$ and how this number varies as the parameters change.
Note that $f(r)$ always has one zero and one turning point for $r<0$.

First, we set $m=1$ (this is equivalent to rescaling the variables 
$\hat{r} = r/m$, $\hat{p} = p/m$, $\hat{a} = a/m^2$).
Zeros of $f$ are then determined (for nonzero $a$) by the zeros
of the $6^{\text{th}}$ order polynomial
\be
p_0(r) = r^6 f(r) = r^6 - 2 r^5 + x r^4 - \frac{2 a x^2 }{5},
\ee
where we write $x = p^2$ for brevity, with $x < 10/9$ from the DEC.
The Sturm method then defines 
\be
p_1(r) = p_0'(r) = 6r^5 - 10 r^4 + 4 x r^3, 
\ee
and involves finding the sequence of remainder polynomials 
$p_{i+1} = - \text{Rem} (p_{i-1},p_i )$. The signs of these polynomials are then
determined as $r \to \pm\infty$, and the difference in the number of sign changes 
at $-\infty$ and $\infty$ gives the number of zeros of $p_0$.

For $r\to\infty$, the signs of the polynomials are
\be
C_i = \text{Sign} \left [ \lim_{r\to\infty} \frac{p_{i}(r)}{r^{6-i}} \right ],
\ee
and the signs are found to be
\be
\beal
C^+_0 &= C^+_1 = C^+_2 = + \,, \qquad C^+_3 = C^+_4 = \text{Sign}[1-x], \\
C^+_5 &= - \text{Sign}[5(25 - 6x)(1 - x) + 27 a (5 - 3x)],\\
C^+_6 &= \text{Sign}[ 1458 a^2 x^2 - 200 x^3 (1 - x) + 5 a ( 3125 - 5625 x + 2700 x^2 - 216 x^3)].
\eeal
\nonumber
\ee
For $r \to -\infty$, the signs of the odd $C^-_i$ are reversed.

For $x<1$, $C_3^+$ and $C_4^+$ are positive, and $C_5^+$ negative, whereas the sign of $C_6^+$ 
depends on the magnitude of $a$.
The sequences are therefore $C_i^+: (+,+,+,+,+ ,-,\pm)$
and $C_i^-: (+,-,+,-,+ ,+,\pm)$. Applying the Sturm method shows that $f$ can have 2 or 4 zeros,
with 4 zeros when $C_6^+<0$.

If $x>1$, $C_3^+$ and $C_4^+$ are negative, while the signs of both $C_5^+$ and $C_6^+$ 
depend non-trivially on the magnitude of $a$, however, 
$C_5^+\geq 0 \Rightarrow C_6^+>0 $. The sequences in this case are 
$C_i^+ : (+,+,+,-,-,+,+)$ or $C_i^+ : (+,+,+,-,-,-,\pm)$ and $C_i^-:(+,-,+,+,-,-,\pm)$ 
or $C_i^-:(+,-,+,+,-,+,\pm)$, again with 2 or 4 zeros, and 4
zeros when $C_6^+<0$.

As noted above, one of the roots of $f$ is always negative, and therefore 
not relevant for the physical domain $r>0$. The remaining positive roots 
determine the causal structure of the spacetime. There can be one or 
three positive roots (counting a repeated root as two), and these 
correspond to physically distinct scenarios. In the case of three positive 
roots, the spacetime is the most similar to the RN spacetime, with the
outermost root marking the event horizon, the second root the Cauchy horizon,
but with an addition of another inner Cauchy horizon due to the $ap^4/r^6$
nonlinear term. With 3 positive roots, $f$ has 2 turning points for $f>0$, and
the coincidence of either of these with the axis corresponds to a type of
extremal limit. If the two inner horizons coincide, this extremal limit has 
no special properties from the point of view of an asymptotic observer, but
the coincidence of the turning point for the outer 2 horizons is a
more conventional extremal limit, with the black hole externally developing
a `throat' and vanishing surface gravity.
If there is only a single positive root, then there is only an
event horizon. In this case, the causal structure is similar to 
the Schwarzschild solution. However in this case, we must be careful 
that the DEC remains satisfied outside the horizon as this case has
the most significant departure from the linear RN solution.
Recall that the DEC requires $r^4_+ \geq a p^2$. Using the 
abbreviation $p^2=x$ and replacing for $a$ in the relation $f(r_+)=0$,
we see that for a horizon on the verge of violating the DEC we have
\be
r_+^2 - 2 r_+ + \frac{9x}{10} \quad \Rightarrow \qquad
r_+ = 1 - \sqrt{1-\frac{9x}{10}},
\ee
(recalling that $r_+\to 0$ as $x \to 0$). Substituting this value into
the expression for $a$ gives 
\be
a \geq \frac{1}{4x} \left ( 1 - \sqrt{1-\frac{9x}{10}} \right )^4,
\ee
for the DEC to be satisfied if we have just one horizon. 

\begin{figure}[h]
\centering
    \includegraphics[width=0.8\textwidth]{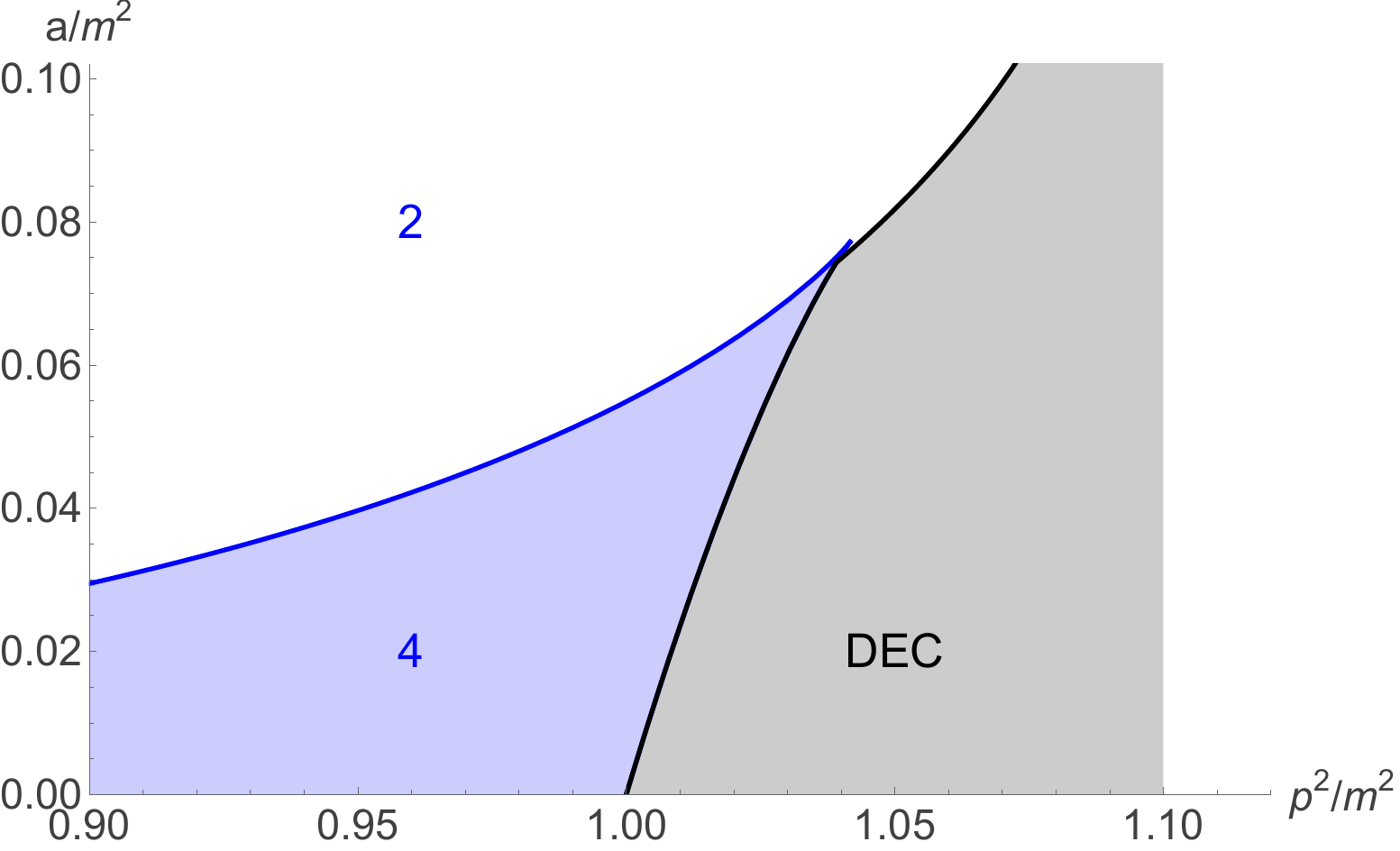}
\caption{A plot of the parameter space indicating the total number of zeros of the lapse function $f(r)$. The blue shaded region is where there are multiple horizons, and the
grey shaded region is excluded due to DEC violation.}
\label{fig:zeros-f}
\end{figure}

Finally, since the number of zeros is determined by the sign of the last
coefficient, $C_6$, solving for $C_6^+ = 0$ gives
\be
a_\pm = \frac{5}{2916 x^2} \left [-3125 + 9x (625 - 300 x + 24 x^2) 
\pm (25-9x)(25-24 x)^{3/2} \right ].
\ee
The vanishing of this coefficient indicates that two of the real roots have 
converged to a repeated root. In this case, the $a_+$ root corresponds 
to the inner Cauchy horizons merging, and the $a_-$ root to the
conventional extremal limit. Figure \ref{fig:zeros-f} shows the 
number of roots of $f$ in parameter space. The shaded region indicates
where we have three horizons in our $r>0$ geometry, 4 zeros in total for $f$,
and the unshaded region is where there is only one horizon (2 zeros of $f$).
The region excluded by DEC violation is also indicated.

\section{Derivation of the effective metrics for \texorpdfstring{$\mathcal{L}(F,G)$}{L(F,G)}}\label{app:characteristics}

In this appendix we present a self-contained derivation of the characteristic 
equation and effective optical metrics for a general nonlinear electrodynamics 
theory with Lagrangian $\mathcal{L}(F,G)$. The approach is based on the principal 
symbol, without invoking Hadamard's discontinuity method.

\medskip
\noindent\textbf{Field equations and linearisation}

The action for nonlinear electrodynamics coupled to Einstein gravity including 
a cosmological constant is
\begin{equation}
\label{eq:action}
S[g_{\mu\nu},A_\mu] = \frac{1}{16\pi G} \int d^4x \, \sqrt{-g} \, 
\bigl(R - 2\Lambda + \mathcal{L}(F,G)\bigr),
\end{equation}
where the Lagrangian $\mathcal{L}(F,G)$ depends on the two independent 
electromagnetic invariants
\begin{equation}
F := F_{\mu\nu}F^{\mu\nu}, \qquad G := F_{\mu\nu}\tilde{F}^{\mu\nu},
\end{equation}
with $\tilde{F}^{\mu\nu} = \tfrac12 \epsilon^{\mu\nu\rho\sigma}F_{\rho\sigma}$ 
the Hodge dual of the field strength. It is convenient to introduce the shorthand
notation for the partial derivatives of the Lagrangian:
\begin{equation}
\mathcal{L}_F := \frac{\partial \mathcal{L}}{\partial F}, \quad
\mathcal{L}_G := \frac{\partial \mathcal{L}}{\partial G}, \quad
\mathcal{L}_{FF} := \frac{\partial^2 \mathcal{L}}{\partial F^2}, \quad
\mathcal{L}_{GG} := \frac{\partial^2 \mathcal{L}}{\partial G^2}, \quad
\mathcal{L}_{FG} := \frac{\partial^2 \mathcal{L}}{\partial F \partial G}.
\end{equation}

Variation of the action with respect to the vector potential $A_\mu$ yields the 
equations of motion
\begin{equation}
\nabla_\mu P^{\mu\nu} = 0, \qquad 
P^{\mu\nu} := \mathcal{L}_F F^{\mu\nu} + \mathcal{L}_G \tilde{F}^{\mu\nu},
\end{equation}
while the definition of the field strength in terms of the potential guarantees 
the Bianchi identity $\nabla_\mu \tilde{F}^{\mu\nu} = 0$. 
Since the analysis of characteristics depends only on the highest-derivative terms,
the gravitational sector $(R - 2\Lambda)$ does not affect the principal symbol; 
the effective metrics derived below are therefore valid for any background 
satisfying the Einstein--NLED field equations.

Linearising around a background field $F_{\mu\nu}$ by writing 
$F_{\mu\nu} \to F_{\mu\nu} + f_{\mu\nu}$ with $|f_{\mu\nu}| \ll 1$, 
the perturbation of the auxiliary tensor takes the form
\begin{equation}
\delta P^{\mu\nu} = P^{\mu\nu\alpha\beta} f_{\alpha\beta} + \mathcal{L}_G \tilde{f}^{\mu\nu},
\end{equation}
where the constitutive tensor $P^{\mu\nu\alpha\beta}$ is given by
\begin{equation}
\label{eq:Pmunualphabetafull}
\begin{aligned}
P^{\mu\nu\alpha\beta} &= \frac{1}{2} \mathcal{L}_F (g^{\mu\alpha}g^{\nu\beta} - g^{\mu\beta}g^{\nu\alpha}) 
+ 2\mathcal{L}_{FF} F^{\mu\nu}F^{\alpha\beta} 
+ 2\mathcal{L}_{GG} \tilde{F}^{\mu\nu}\tilde{F}^{\alpha\beta} \\
&\quad + 2\mathcal{L}_{FG}(F^{\mu\nu}\tilde{F}^{\alpha\beta} + \tilde{F}^{\mu\nu}F^{\alpha\beta}).
\end{aligned}
\end{equation}
The term proportional to $\mathcal{L}_G$ does not contribute to the principal 
symbol since $\xi_\mu \tilde{f}^{\mu\nu} = 0$ follows from the Bianchi identity.

\medskip
\noindent\textbf{Principal symbol and reduced system}

The principal symbol is obtained by replacing derivatives with the wave covector, 
$\nabla_\mu \mapsto \xi_\mu$. The full principal symbol of the linearised theory is
\begin{equation}
\sigma_L(\xi)^{\nu}{}_{\alpha\beta} = \xi_\mu P^{\mu\nu}{}_{\alpha\beta}.
\end{equation}
Writing the field perturbation in the form $f_{\alpha\beta} = \xi_\alpha a_\beta - \xi_\beta a_\alpha$, 
the characteristic condition $\xi_\mu \delta P^{\mu\nu} = 0$ becomes
\begin{equation}
2 P^{\mu\nu\alpha\beta} \xi_\mu \xi_\alpha a_\beta = 0.
\end{equation}
Contracting with the metric defines the reduced principal symbol
\begin{equation}
\label{eq:McM}
\mathcal M^\nu{}_\beta := 2\, \xi_\mu \, P^{\mu\nu\alpha\gamma} \, \xi_\alpha \, g_{\gamma\beta}
= 2\,\sigma_L(\xi)^{\nu}{}_{\alpha\beta}\,\xi^\alpha.
\end{equation}
The characteristic equation is then $\mathcal{M}^\nu{}_\beta a^\beta = 0$, and 
nontrivial polarisations exist only when $\ker\mathcal{M} \neq \{0\}$ on the 
three-dimensional subspace transverse to $\xi_\mu$. Gauge invariance guarantees 
$\mathcal{M}^\nu{}_\beta \xi^\beta = 0$, so physical polarisations lie in the 
quotient space $\mathcal{P}_\xi = \{a_\mu : \xi^\mu a_\mu = 0\} / \langle\xi_\mu\rangle$.

\medskip
\noindent\textbf{Explicit form of the reduced symbol}

Substituting the constitutive tensor \eqref{eq:Pmunualphabetafull} into 
\eqref{eq:McM} yields the explicit reduced symbol
\begin{align}
\label{eq:Mfull}
\mathcal{M}^\nu{}_\beta
&= \mathcal{L}_F (\xi^2 \delta^\nu_\beta - \xi^\nu \xi_\beta) 
+ 4\mathcal{L}_{FF} (\xi_\mu F^{\mu\nu})(\xi_\alpha F^{\alpha}{}_\beta) \\
&\quad + 4\mathcal{L}_{GG} (\xi_\mu \tilde{F}^{\mu\nu})(\xi_\alpha \tilde{F}^{\alpha}{}_\beta) 
+ 4\mathcal{L}_{FG}\Big[(\xi_\mu F^{\mu\nu})(\xi_\alpha \tilde{F}^{\alpha}{}_\beta) 
+ (\xi_\mu \tilde{F}^{\mu\nu})(\xi_\alpha F^{\alpha}{}_\beta)\Big], \nonumber
\end{align}
where $\xi^2 = g^{\mu\nu}\xi_\mu\xi_\nu$. Defining the transverse vectors
\begin{equation}
v^\mu := \xi_\alpha F^{\alpha\mu}, \qquad w^\mu := \xi_\alpha \tilde{F}^{\alpha\mu},
\end{equation}
which satisfy $\xi_\mu v^\mu = \xi_\mu w^\mu = 0$ by antisymmetry, the symbol 
takes the compact form
\begin{equation}
\label{eq:Mcompact}
\mathcal{M}^\nu{}_\beta = \mathcal{L}_F(\xi^2 \delta^\nu_\beta - \xi^\nu \xi_\beta)
+ 4\mathcal{L}_{FF} v^\nu v_\beta 
+ 4\mathcal{L}_{GG} w^\nu w_\beta
+ 4\mathcal{L}_{FG}(v^\nu w_\beta + w^\nu v_\beta).
\end{equation}

\medskip
\noindent\textbf{Polarisation decomposition}

Any transverse vector $a^\mu$ (with $\xi_\mu a^\mu = 0$) admits the decomposition
\begin{equation}
a^\mu = \alpha v^\mu + \beta w^\mu + a_\perp^\mu,
\end{equation}
where $a_\perp^\mu$ is orthogonal to $v^\mu$, $w^\mu$, and $\xi_\mu$. 
Applying $\mathcal{M}^\nu{}_\beta$ to $a_\perp^\beta$ yields
\begin{equation}
\mathcal{M}^\nu{}_\beta a_\perp^\beta = \mathcal{L}_F \xi^2 a_\perp^\nu,
\end{equation}
since all contractions with $v_\beta$ and $w_\beta$ vanish. 
This gives the condition
\begin{equation}
\label{eq:orthogonal}
\mathcal{L}_F \xi^2 = 0 \quad \Longrightarrow \quad g^{\mu\nu}\xi_\mu\xi_\nu = 0,
\end{equation}
corresponding to propagation along the background spacetime light cones.

For the parallel subspace spanned by $\{v^\mu, w^\mu\}$, writing 
$a_\parallel^\mu = \alpha v^\mu + \beta w^\mu$, the action of $\mathcal{M}$ 
reduces to a $2\times 2$ matrix:
\begin{equation}
\label{eq:Mmatrix}
\mathbf{M} = \mathcal{L}_F \xi^2 \mathbb{I}_2 + 4\mathbf{L}\mathcal{T},
\end{equation}
with
\begin{equation}
\mathbf{L} := 
\begin{pmatrix}
\mathcal{L}_{FF} & \mathcal{L}_{FG} \\
\mathcal{L}_{FG} & \mathcal{L}_{GG}
\end{pmatrix},
\qquad
\mathcal{T} :=
\begin{pmatrix}
v\!\cdot\!v & v\!\cdot\!w \\
v\!\cdot\!w & w\!\cdot\!w
\end{pmatrix}.
\end{equation}

\medskip
\noindent\textbf{Characteristic determinant and effective metrics}\label{app:sec:char_det}

The characteristic condition on the parallel subspace is $\det\mathbf{M} = 0$, 
which yields a quadratic equation in $\mathcal{L}_F\xi^2$:
\begin{equation}
\det\mathbf{M} = (\mathcal{L}_F\xi^2)^2 + \mathcal{L}_F\xi^2 \tr(4\mathbf{L}\mathcal{T}) 
+ \det(4\mathbf{L}\mathcal{T}) = 0.
\end{equation}
Factorising, we obtain
\begin{equation}
\label{eq:factorization}
\det\mathbf{M} = (\mathcal{L}_F\xi^2 - 4\mu_+)(\mathcal{L}_F\xi^2 - 4\mu_-) = 0,
\end{equation}
where $\mu_\pm$ are the eigenvalues of $-\mathbf{L}\mathcal{T}$:
\begin{equation}
\label{eq:mu}
\mu_\pm = \frac{1}{2}\!\left(-\tr(\mathbf{L}\mathcal{T}) 
\pm \sqrt{(\tr(\mathbf{L}\mathcal{T}))^2 - 4\det(\mathbf{L}\mathcal{T})}\right).
\end{equation}
Each factor defines a quadratic form
\begin{equation}
\mathcal{L}_F \xi^2 - 4\mu_\pm =: \mathcal{G}^{\mu\nu}_\pm \xi_\mu \xi_\nu = 0,
\end{equation}
from which we read off the two effective optical metrics. 
Explicit expressions for $\mathcal{G}^{\mu\nu}_\pm$ can be written in closed form 
for specific subclasses; the most general case leads to a Fresnel-type equation 
whose explicit metric coefficients involve the background invariants $F$, $G$ and 
the derivatives of $\mathcal{L}$.

\medskip
\noindent\textbf{Special case: \texorpdfstring{$\mathcal{L}_{FG}=0$}{L\_FG=0} and \texorpdfstring{$G=0$}{G=0}}\label{app:sec:special_case}

When $\mathcal{L}_{FG} = 0$ and the background has $G = 0$, the matrices 
$\mathbf{L}$ and $\mathcal{T}$ become diagonal, yielding
\begin{align}
\mu_+ &= -\mathcal{L}_{GG} w^2, \\
\mu_- &= -\mathcal{L}_{FF} v^2.
\end{align}
The corresponding effective metrics are
\begin{align}
\label{eq:Gplus}
\mathcal{G}^{\mu\nu}_+ &= \mathcal{L}_F g^{\mu\nu} 
+ 4\mathcal{L}_{GG} \, g_{\rho\sigma} \tilde{F}^{\mu\rho} \tilde{F}^{\nu\sigma}, \\[0.5em]
\label{eq:Gminus}
\mathcal{G}^{\mu\nu}_- &= \mathcal{L}_F g^{\mu\nu} 
+ 4\mathcal{L}_{FF} \, g_{\rho\sigma} F^{\mu\rho} F^{\nu\sigma}.
\end{align}

The full characteristic determinant, including the orthogonal polarisation 
contribution \eqref{eq:orthogonal}, factorises as
\begin{equation}
\det\nolimits_\perp \mathcal{M} \propto (g^{\mu\nu}\xi_\mu\xi_\nu)\,
(\mathcal{G}^{\mu\nu}_+ \xi_\mu\xi_\nu)\,
(\mathcal{G}^{\mu\nu}_- \xi_\mu\xi_\nu) = 0.
\end{equation}

\medskip
\noindent\textbf{Additional subcases.}

In the same way it is straightforward to find the following limits:

\begin{itemize}

\item \textbf{Maxwell limit} ($\mathcal{L} = -\tfrac14 F$): 
$\mathcal{G}^{\mu\nu}_\pm = g^{\mu\nu}$, recovering the spacetime null cones.

\item \textbf{$\mathcal{L}(F)$ alone} ($\mathcal{L}_{G} = \mathcal{L}_{GG} = \mathcal{L}_{FG} = 0$): 
Then $\mu_+ = 0$, $\mu_- = -\mathcal{L}_{FF} v^2$, giving
\[
\mathcal{G}^{\mu\nu}_+ = \mathcal{L}_F g^{\mu\nu}, \qquad
\mathcal{G}^{\mu\nu}_- = \mathcal{L}_F g^{\mu\nu} + 4\mathcal{L}_{FF} \, 
g_{\rho\sigma} F^{\mu\rho} F^{\nu\sigma}.
\]
No birefringence occurs.

\item \textbf{$\mathcal{L}(G)$ alone} ($\mathcal{L}_F = \mathcal{L}_{FF} = \mathcal{L}_{FG} = 0$): 
Then $\mu_- = 0$, $\mu_+ = -\mathcal{L}_{GG} w^2$, giving
\[
\mathcal{G}^{\mu\nu}_- = 0, \qquad
\mathcal{G}^{\mu\nu}_+ = 4\mathcal{L}_{GG} \,
g_{\rho\sigma} \tilde{F}^{\mu\rho} \tilde{F}^{\nu\sigma}.
\]
The reduced symbol reduces to $\mathcal{M}^\nu{}_\beta = 4\mathcal{L}_{GG} w^\nu w_\beta$, 
which has rank $\le 1$, and the characteristic determinant vanishes identically. 
This theory is physically distinct from the previous cases: it possesses no 
Maxwell limit and has a degenerate Cauchy problem, signalling that pure 
$\mathcal{L}(G)$ theories are not viable as stand-alone models.
\end{itemize}

\bibliographystyle{apsrev4-2}
\bibliography{bib}

\end{document}